\documentclass{aa}
\usepackage{psfig}

\newcommand{\ms}{$M_{\odot}$}

\newcommand{\mss}{$M_{\odot}$}
\newcommand{\eg}{\rm e.g.\ }
\newcommand{\etal}{\rm et al.\ }
\newcommand{\ie}{\rm i.e.\/ }

\newcommand{\cf}{cf.\/ }
\newcommand{\bull}{}
\newcommand{\blok}{}

\newcommand{\lsbfil}[1]{./#1}

\def\la{\mathrel{\mathchoice {\vcenter{\offinterlineskip\halign{\hfil
$\displaystyle##$\hfil\cr<\cr\sim\cr}}}
{\vcenter{\offinterlineskip\halign{\hfil$\textstyle##$\hfil\cr
<\cr\sim\cr}}}
{\vcenter{\offinterlineskip\halign{\hfil$\scriptstyle##$\hfil\cr
<\cr\sim\cr}}}
{\vcenter{\offinterlineskip\halign{\hfil$\scriptscriptstyle##$\hfil\cr
<\cr\sim\cr}}}}}

\def\ga{\mathrel{\mathchoice {\vcenter{\offinterlineskip\halign{\hfil
$\displaystyle##$\hfil\cr>\cr\sim\cr}}}
{\vcenter{\offinterlineskip\halign{\hfil$\textstyle##$\hfil\cr
>\cr\sim\cr}}}
{\vcenter{\offinterlineskip\halign{\hfil$\scriptstyle##$\hfil\cr
>\cr\sim\cr}}}
{\vcenter{\offinterlineskip\halign{\hfil$\scriptscriptstyle##$\hfil\cr
>\cr\sim\cr}}}}}

\begin{document}

\thesaurus{03(11.05.2; 11.06.2; 11.06.1; 11.19.2)}

\title{The evolution of the stellar populations in 
low surface brightness galaxies}

\author{L.B.~van den Hoek\inst{1}, 
W.J.G.~de Blok\inst{2,3}\thanks{Bolton Fellow},
J.M.~van der Hulst\inst{2}, and T.~de Jong\inst{1,4} }

\institute{Astronomical Institute 'Anton Pannekoek',
   Kruislaan 403, NL-1098 SJ Amsterdam, 
   The Netherlands
\and 
   Kapteyn Astronomical Institute, 
   P.O.~Box 800, NL-9700 AV Groningen, The Netherlands
\and
   Australia Telescope National Facility, P.O.~Box 76, Epping NSW 1710, Australia
\and
   Space Research Organisation of the 
   Netherlands, Sorbonnelaan 2,  NL-3584 CA Utrecht, The Netherlands}

\date{Received date; accepted date}

\offprints{W.J.G. de Blok, {\tt edeblok@atnf.csiro.au}}

\titlerunning{Evolution of LSB galaxies}
\authorrunning{L.B. van den Hoek \etal}
\maketitle

\begin{abstract} 
We investigate the star formation history and chemical evolution of
low surface brightness (LSB) disk galaxies by modelling their observed
spectro-photometric and chemical properties using a galactic chemical
{ and photometric} evolution model incorporating a detailed metallicity dependent set of
stellar input data.  
For a large fraction of the LSB galaxies in our sample, observed
properties are best explained by models incorporating an exponentially
decreasing global star formation rate (SFR) ending at a present-day
gas fraction $M_{\rm gas}/(M_{\rm gas}+M_{\rm stars}) = 0.5$ for a
galaxy age of 14 Gyr.  For some galaxies small amplitude star
formation bursts are required to explain the contribution of the young
(5-50 Myr old) stellar population to the galaxy integrated luminosity.
This suggests that star formation has proceeded in a stochastic
manner.  

The presence of an old stellar population in many late-type LSB
galaxies suggests that LSB galaxies roughly follow the same
evolutionary history as HSB galaxies, {\em except at a much lower
rate}. In particular, our results imply that LSB galaxies do not form
late, nor have a delayed onset of star formation, but simply evolve
slowly.


\end{abstract}

\section{Introduction}

Deep searches for field galaxies in the local universe have revealed
the existence of a large number of galaxies with low surface
brightnesses that  are hard to detect against the night sky (see
the reviews by Impey \& Bothun 1997 and Bothun, Impey \& McGaugh
1997).  Many different kinds of LSB galaxies have so far been found,
including giant LSB galaxies (e.g. Sprayberry et al.\ 1995) and red
LSB galaxies (O'Neil et al.\ 1997), but the most common type seems to
be ``blue LSB galaxies'': late-type, disk-dominated spirals with
central surface brightnesses $\mu_0(B) \ga 23$ mag arcsec$^{-2}$.

In this paper, we will concentrate on these late-type LSB spirals.
Observations show that they are neither exclusively dwarf systems nor just the
faded counterparts of ``normal'' high surface brightness (HSB) spirals
(de Blok \etal 1996, hereafter dB96).  In many cases, LSB galaxies
follow the trends in galaxy properties found along the Hubble sequence
towards very late types.  These trends include increasingly blue
colours (\eg R\"onnback 1993; McGaugh \& Bothun 1994; de Blok et al.\
1995 [hereafter dB95]; Bell et al.\ 1999), decreasing oxygen
abundances in the gas (\eg McGaugh 1994; R\"onnback \& Bergvall 1995;
de Blok \& van der Hulst 1998a), and decreasing H{\sc i} surface
densities (from type Sc onwards; \eg van der Hulst \etal 1993,
hereafter vdH93).  Despite the low gas densities, LSB spirals rank
among the most gas-rich disk galaxies at a given total luminosity as
their H{\sc i} disks in general are extended (Zwaan \etal 1995; dB96).
$\bull $\bull $\bull$ The fact that LSB galaxies still have large
reservoirs of gas together with their low abundances suggest that the
amount of star formation in the past cannot have been very large
(vdH93; Van Zee \etal 1997). Clearly, LSB galaxies are not the faded
remnants of HSB spirals.  $\bull \bull
\bull$

The unevolved nature of LSB spirals can be interpreted in various
ways.  For instance, LSB spirals could be systems whose stellar
population is young and in which the main phase of star formation is
still to occur.  Alternatively, the stellar population in LSB spirals
could be similar to those in HSB galaxies but with a young population
dominating the luminosity.  This has been explored and modelled in
studies by e.g.\  Knezek (1993), Jimenez et al.\ (1998), dB96, Gerritsen
\& de Blok (1999) and more recently in two papers by Bell et al.\
(1999) and Bell \& de Jong (1999). These studies all indicate a
scenario in which LSB galaxies are unevolved systems with low surface
densities, low metallicities and a constant or increasing star
formation rate (cf.\ Padoan et al.\ 1997, however, for a different opinion).

In this paper we address these various scenarios for the evolution of
LSB galaxies by detailed modelling of their spectro-photometric and
chemical properties.  Model results are compared directly with the
observed colours, gas phase abundances, gas contents, and current star
formation rates of LSB galaxies. Using simple model star formarion
histories we confirm the results of previous studies and show that the
ratio of young stars to old stars in LSB galaxies is larger than
typically found in HSB galaxies.

In reality, however, star formation is unlikely to proceed as smoothly
as the various population synthesis models suggest. Gerritsen \& de
Blok (1999) point out that small surges in the star formation rate are likely
to be very important in determining the observed colours of LSB
galaxies.  In this paper we will pay special attention to the influence of
small bursts of star formation on the optical colours, and show that
they can have a significant influence on the inferred properties of
LSB galaxies.

This paper is organized as follows.  We briefly compare observational
data of LSB galaxies with those of spirals and dwarf galaxies in
Sect.\ 2.  In Sect.\ 3, we describe the ingredients of the galactic
evolution model we used.  Some general properties of the model are
presented in Sect.\ 4.  The comparison with observational data is
presented and discussed in Sect.\ 5.  The impact of small amplitude
star formation bursts in LSB galaxies is investigated in Sect.\ 6 and
predicted star formation rates are compared with the observations in
Sect.\ 7.  A short discussion of where LSB galaxies fit in the grand
scheme of { galaxy formation and evolution} is presented in Sect.\
8. A short summary is given in Sect.\ 9.

\section{Observational characteristics of LSB galaxies}

\subsection{Sample selection}

We use the sample described in dB96. This consists of
24 late-type LSB galaxies (inclinations up to $\sim 60^{\circ}$),
taken from the lists by Schombert \etal (1992) and the UGC (Nilson
1973). The sample is representative for the LSB galaxies generally
found in the field.  We selected a subsample of 16 LSB galaxies for
which good data are available.  For these systems, optical data have
been taken from dB95, H{\sc i} data from dB96, and abundance data from
McGaugh \& Bothun (1994) and de Blok \& van der Hulst (1998a).

The galaxy identification and Johnson $UBV$ and Kron-Cousins $RI$ are
given in columns (1) to (6) in Table~1 (a Hubble constant of $H_0 =
100$ km s$^{-1}$ Mpc$^{-1}$ was used). \blok Column (7) lists the
$B$-band luminosities; column (8) the H{\sc i}-mass-to-light ratio;
column (9) and (10) the H{\sc i} and dynamical masses respectively.
The gas fraction $\mu_{\rm rot}$ in column (11) is given by $M_{\rm
gas}$/$(M_{\rm gas}+ M_{*,{\rm max}})$ where $M_{*,{\rm max}}$ denotes
the mass of the stellar component obtained from maximum disk fitting
of the rotation curve (dB96) (see Sect.\ 2.6).  The mean [O/H]
abundances are listed in column (12).  \blok

\begin{table*}[tbh]
\caption[]{Observational data on LSB galaxies (de Blok \etal 1995, 1996)}
\begin{tabular}{llllllllllll}
\hline
(1) & (2) & (3) & (4) & (5) & (6) & (7) & (8) & (9) & (10) & (11) & (12) \\
\noalign{\vskip 2pt}
Name  & $U$ &  $B$ &  $V$  &  $R$  &  $I$ &  $\log L_{B}$ & 
${M_{\rm HI}}\over{L_{B}}$ & $\log M_{\rm HI}$ & $\log M_{\rm dyn}$ & $\mu_{\rm rot}$ & [O/H] \\
\noalign{\vskip 2pt}
\hline
F561-1  & $-$17.4 & $-$17.2 & $-$17.8 & $-$18.0 & $-$18.4 & 9.08 & 0.7 & 
8.91 & 9.66  & 0.59 &   $-$0.87 \\
F563-1  & $-$16.6 & $-$16.7 & $-$17.4 & $-$17.6 & $-$16.4\dag & 8.87 & 2.0 & 
9.18 & 10.58 & 0.19 &  $-$1.46 \\
F563-V1 & $-$15.7 & $-$15.7 & $-$16.3 & $-$16.6\dag & $-$16.9 & 8.48 & 0.9 & 
8.45 & 9.00  & 0.59 &  $-$1.05 \\
F564-V3 &    *   & $-$11.8 & $-$12.4 & $-$12.6 &   *    & 6.91 & 1.6 & 
7.11 & 8.80  & * &    *    \\
F565-V2 &     *  & $-$14.8 & $-$15.3 & $-$15.6 &    *   & 8.11 & 2.6 & 
8.53 & 9.58  & 0.33 &     *  \\
F567-2  & $-$17.0 & $-$16.8 & $-$17.4 & $-$17.4 & $-$17.7 & 8.91 & 1.5 & 
9.08 & 9.90  & 0.49 &     *   \\
F568-1  & $-$17.7 & $-$17.5 & $-$18.1 & $-$18.3 & $-$18.8 & 9.20 & 1.4 & 
9.34 & 10.56 & 0.32 &    $-$0.98  \\
F568-3  & $-$17.8 & $-$17.7 & $-$18.2 & $-$18.5 & $-$19.0 & 9.28 & 0.8 & 
9.20 & 10.63 & 0.45 &  $-$0.92 \\
F568-V1 & $-$17.4 & $-$17.3 & $-$17.8 & $-$18.0 & $-$18.5 & 9.11 & 1.1 & 
9.15 & 10.71\dag & 0.53 &  $-$0.99 \\
F571-5  & $-$16.6 & $-$16.5 & $-$16.9 & $-$17.1 &   *    & 8.79 & * & 
9.99\dag & 10.45 & * &  $-$1.53 \\
F571-V1 &   *    & $-$16.4 & $-$16.9 & $-$17.3 &    *   & 8.75 & 1.2 & 
8.82 & 10.15 & 0.33 &  $-$0.94 \\
F574-2  &  *     & $-$17.0 & $-$17.7 & $-$17.8 &   *    & 8.99 & 0.9 & 
8.97 & 9.51 & 0.66 &    *    \\
F577-V1 & $-$17.9 & $-$17.6 & $-$18.0 & $-$18.1 &   *    & 9.23 & 0.8 & 
9.15 & 9.20\dag & * &   *   \\
U0128   & $-$18.5 & $-$18.2 & $-$18.7 & $-$18.9 & $-$19.3 & 9.48 & 1.2 & 
9.56 & 10.86 & 0.39 &    *    \\
U0628   & $-$18.5 & $-$18.5 & $-$19.1 & $-$19.3 & $-$19.7 &  9.59 & * &  
*    & 10.69 & * &    *    \\
U1230   & $-$18.9 & $-$17.7 & $-$18.2 & $-$18.5 & $-$18.9 & 9.28 & 1.7 & 
9.50 &  10.80   & 0.60 &  $-$0.76 \\ \hline
Typ. Dwarf & $-$17.5 & $-$17 & $-$17.5 & $-$18 & $-$18.5 & 9.0 & 1.0 & 
9.0 & 9.0   & 0.7 &  $-$0.6$\pm$0.4 \\
Typ. LSBG &  $-$18    & $-$17.5 & $-$18 & $-$18 & $-$18.5 & 9.2 & 1.3 & 
9.3 & 10.0 & 0.5 &  $-$0.6$\pm$0.4 \\
Typ. HSBG &  $-$20    & $-$19.5 & $-$20 & $-$20.5 & $-$21 & 10.0 & 0.4 & 
9.6 & 11.0 & 0.1 & $\sim$0.0 
\\ \hline \\
\end{tabular}
\parbox{12.cm}{{\bf *} not available, {\bf \dag} uncertain }
\end{table*}

\begin{figure*}[p] \vspace{0.cm}
{\psfig{figure=\lsbfil{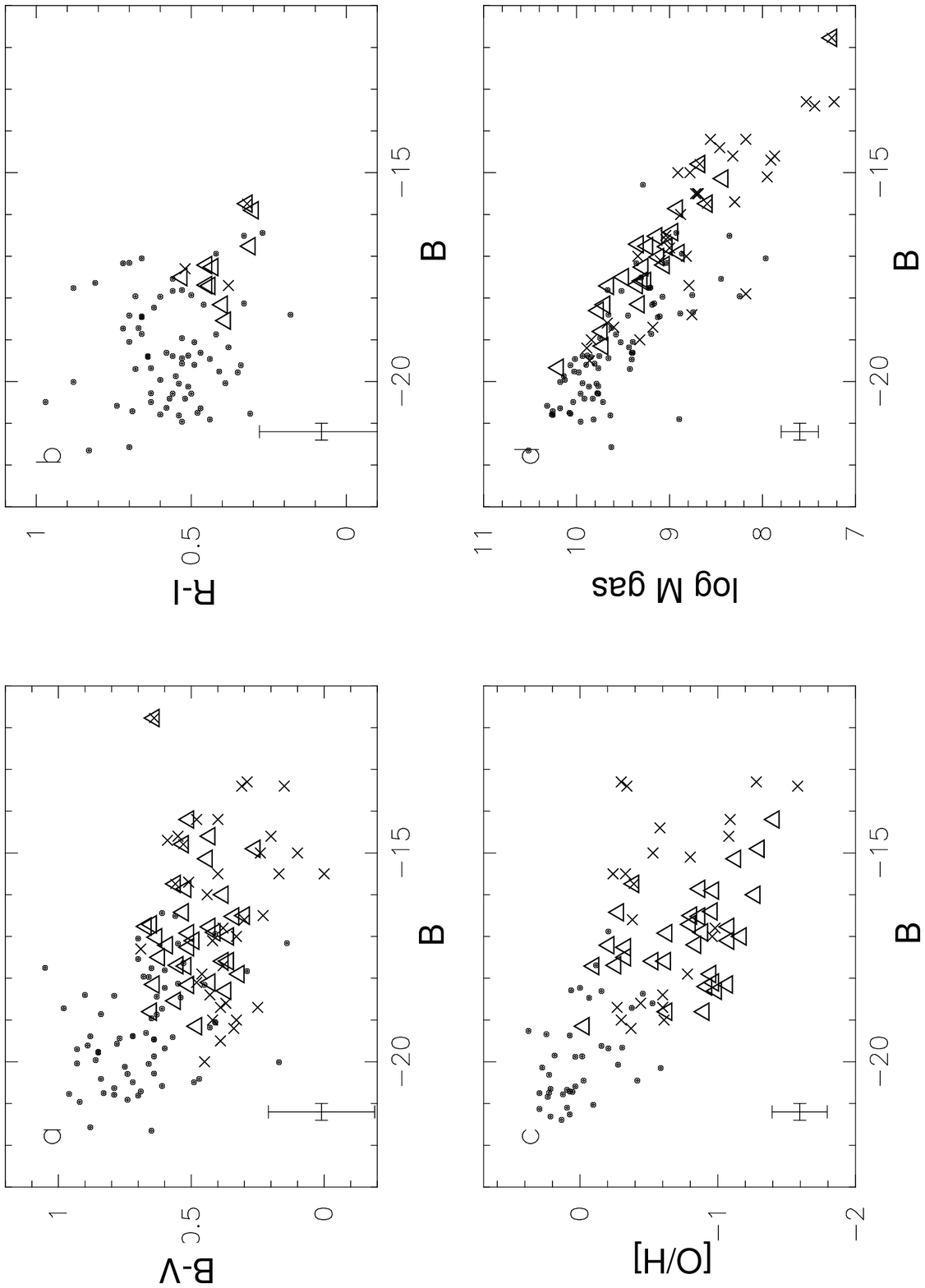},height=11.cm,width=16.cm,angle=270.}}
\vspace{0.5cm} 
{\psfig{figure=\lsbfil{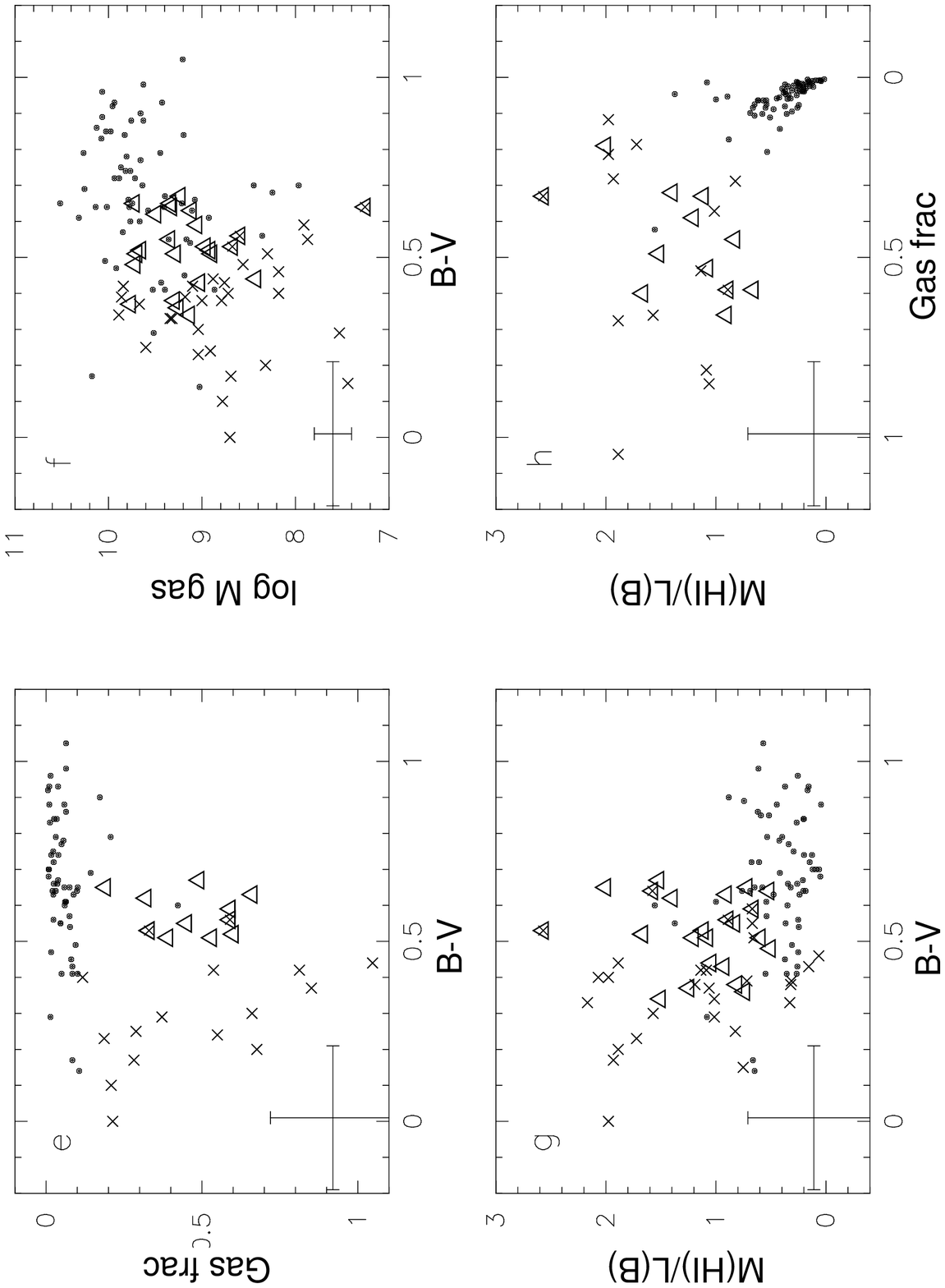},height=11.cm,width=16.cm,angle=270.}}
\caption[]{Observational data on LSB galaxies compared to that for
normal spirals and dwarf galaxies. Symbols refer to LSB galaxies
(triangles; data from de Blok \etal 1995, 1996), normal face-on
spirals (full dots; de Jong \& van der Kruit 1994), and dwarf galaxies
(crosses; Mellise \& Israel 1994). LSB galaxies which are probably
dwarf systems are indicated by triangles with crosses
overlayed. Typical error bars are indicated in the bottom left of each
panel.}
\end{figure*}

\subsection{Magnitudes and colours}

We compare in Fig.~1$a$ and $b$ magnitudes and broadband colours of
LSB galaxies to those of \blok late-type \blok HSB face-on spirals (de
Jong \& van der Kruit 1994) and dwarf galaxies (Melisse \& Israel
1994; Gallagher \& Hunter 1986, 1987).  The usual distinction between
HSB and LSB galaxies is purely an artificial one since normal galaxies
along the Hubble sequence show a {\it continuous} range in central
surface brightness, \ie ranging from values around the Freeman value
for early type systems to the very faint values observed for the
late-type LSB galaxies in our sample (\eg de Jong 1995; dB95).  For
the purpose of this paper we will continue to use the distinction as a
shorthand, where ``HSB galaxy'' should be interpreted as ``typical
Hubble sequence Sc galaxy.''  { Fig.\ 1$a$ and $b$ show that} LSB
galaxies are { generally} bluer and fainter than their HSB counterparts
(\eg McGaugh 1992; vdH93; dB95).

\subsection{Abundances}

Estimates of the ISM abundances in LSB galaxies predominantly rely on
abundance determinations of their H{\sc ii} regions. 
Oxygen abundances of H{\sc ii} regions listed in Table~1 are taken
from McGaugh (1994) and de Blok \& van der Hulst (1998a).

We assume that the H{\sc ii}-region abundances on average are a
reasonable indicator of the ISM abundances within a given LSB spiral.
$\bull $\bull $\bull$ The intrinsic scatter in [O/H] among different
H{\sc ii} regions within a given LSB galaxy is usually of the order
0.2 dex (\eg McGaugh 1994). $\bull \bull \bull$ Bell \& de Jong (1999)
find that the gas metal abundances in LSB galaxies generally trace the
stellar metallicities quite well, with the stars being $\sim 0.4$ dex
more metal-poor.

In Fig.\ 1c we compare mean [O/H] abundances of H{\sc ii} regions in
LSB galaxies with those in HSB spirals (Zaritsky \etal 1994) and dwarf
galaxies (Melisse \& Israel 1994; Gallagher \& Hunter 1986, 1987).  On
average, LSB galaxies generally follow the correlation between the
characteristic gas-phase abundance and luminosity as found for HSB
spirals (Zaritsky \etal 1994), but with a large scatter. The range in
[O/H] at a given $B$ magnitude is nearly one dex.  $\bull $\bull
$\bull$ This scatter is probably related to evolutionary differences
among the LSB galaxies of a given $B$ magnitude (\eg in the ratio of
old to young stellar populations) and/or the H{\sc ii} regions they
contain. $\bull \bull \bull$

\subsection{Extinction}


Observational studies by (amongst others) Bosma \etal (1992), Byun
(1992), Tully \& Verheijen (1997) show that late-type spirals appear
transparent throughout their disks.  This supports the idea that
face-on extinction in LSB galaxies is relatively low, \ie $E_{B-V} \la
0.1$ (\eg McGaugh 1994).  This is consistent with findings by Tully \&
Verheijen (1998), who studied a large sample of galaxies in the UMa
cluster and found the LSB galaxies in their sample to be transparent,
whereas non-negligible extinction was present in the HSB galaxies.

In this study we will ignore the possible effects of dust on the colours and magnitudes. See Sect.\ 5.3 for a discussion.

\subsection{Gas masses}

We compare in Fig.~1d the observed atomic gas masses $M_{\rm g}
\sim 1.4 M_{\rm HI}$ (corrected for helium) of LSB galaxies with
those of HSB galaxies and dwarfs. Strictly speaking we should also
take the molecular gas mass into account.  However, so far no
CO-emission has been detected in late-type LSB galaxies (Schombert
\etal 1990; de Blok \& van der Hulst 1998b, but see Knezek 1993 for a discussion on giant LSB galaxies).
The amount of molecular gas in LSB galaxies is probably small, even
though relatively high CO/H$_{2}$ conversion factors may apply in
these low metallicity galaxies (Wilson 1995, Mihos et al.\ 1999).
$\bull $\bull $\bull$ We conclude that, on average, LSB galaxies are
considerably more gas-rich (up to a factor $\sim 3$) than HSB galaxies
{\em of the same $B$ magnitude} (see also Fig.\ 1g). $\bull \bull
\bull$

\subsection{Gas fractions and mass-to-light ratios}

We show in Figs.~1g and h the distribution of the mass-to-light ratio
$M_{\rm HI}/L_{B}$ {\it vs} $(B-V)$ and {\it vs} $\mu_{rot}$,
respectively.  LSB galaxies exhibit considerably higher $M_{\rm HI}/
L_{B}$ ratios than HSB spirals (\cf Table~1 and McGaugh \& de Blok
1997).

We would however like to know the ratio of gas mass to stellar
mass, i.e.\ the gas fraction $\mu_{1}
\equiv M_{\rm gas}$ / $(M_{\rm gas}+M_{\rm stars}$ of a galaxy. This
involves the conversion of the observed galaxy luminosity to stellar
mass. However, the mass-to-light ratio of the underlying stellar
population is generally not well known and, in fact, is an important
quantity to determine.  Assuming a fixed value of the mass-to-light
ratio will however introduce artificial trends since this ratio is
expected to vary among galaxies having different star formation
histories (e.g.\ Larson \& Tinsley 1978).

As a partial solution we have determined the gas fraction $\mu_{\rm
rot} \equiv M_{\rm gas} / (M_{\rm gas}+ M_{*, {\rm max}})$ where
$M_{*,{\rm max}}$ denotes the mass of the stellar component obtained
from maximum disk fitting of the rotation curve (dB96).  This method
likely overestimates the contribution of the luminous stellar disk to
the observed total mass distribution by about a factor of 2 (\eg
Kuijken \& Gilmore 1989; Bottema 1995) and, therefore, also provides a
lower limit to the actual gas fraction. { In summary, while $\mu_1$
is a {\it model} quantity, $\mu_{\rm rot}$ is determined from {\it observations}.}

Figs.\ 1e and 1f show the distribution of the present-day gas fraction
$\mu_{\rm rot}$ and total gas mass {\it vs} $(B-V)$. It can be seen
that LSB galaxies and dwarfs exhibit larger gas fractions than HSB
spirals. Taking into account the factor of $\sim 2$ overestimate,
values for LSB galaxies are typically $\mu_{1} \ga 0.5$ and $\mu_{1}
\sim 0.1$ for HSB spirals. 


We conclude that the available evidence supports of the view that LSB
galaxies are relatively unevolved systems compared to HSB spirals.
This agrees well with the fact that LSB galaxies usually display
properties intermediate to those of HSB spirals and dwarf galaxies.

\section{Model description and assumptions}

We here summarise the galactic evolution model used (for a more
extensive description see van den Hoek 1997, hereafter vdH97).
We concentrate on the stellar contribution to the total galaxy
luminosity in a given passband (other contributions are
neglected). For a given star formation history (SFH), we compute the
chemical enrichment of a model galaxy by successive generations of
evolving stars.  To derive the stellar luminosity in a given passband
at a given age, we use a metallicity dependent set of theoretical
stellar isochrones as well as a library of spectro-photometric
data. 

\subsection{Chemical evolution model}

We start from a model galaxy initially devoid of stars.  We follow the
chemical enrichment during its evolution assuming stars to be formed
according to a given star formation rate (SFR) and initial mass
function (\eg a power law IMF: d$N$/d$m \equiv M(m)
\propto m^{\gamma}$).  Both stellar and interstellar abundances 
as a function of galactic evolution time $t_{ev}$ are computed assuming
that the stellar ejecta are returned and homogeneously mixed to the
ISM at the end of { the individual stellar  lifetimes} (\ie relaxing the instantaneous
recycling approximation; see Searle \& Sargent 1972). A description of
the set of galactic chemical evolution equations used can be found in
\eg Tinsley (1980) and Twarog (1980), see also vdH97. This model is a
closed-box model.

We follow the stellar enrichment of the star forming galaxy in terms
of the characteristic element contributions of Asymptotic Giant Branch
(AGB) stars, SNII and SNIa. This treatment is justified by the
specific abundance patterns observed within the ejecta of these
stellar groups (see \eg Trimble 1991; Russell \& Dopita 1992).  A
detailed description of the metallicity dependent stellar lifetimes,
element yields, and remnant masses is given by vdH97 and
van den Hoek \& Groenewegen (1997).  We compute the abundances of H,
He, O, and Fe, as well as the heavy element integrated metal-abundance
$Z$ (for elements more massive than helium), during the evolution of
the model galaxy. Both the SFH, IMF and resulting element abundances
as a function of galactic evolution time are used as input for the
spectro-photometric evolution model described below.

Boundary conditions to the chemical evolution model are the galaxy
total mass $M_{\rm tot}$, its evolution time $t_{\rm ev}$, and the
initial gas abundances.  Note that solutions of the galactic chemical
evolution equations are independent of the {\em ratio} of the SFR
normalisation and $M_{\rm tot}$.  Primordial helium and hydrogen
abundances are adopted as $Y_{\rm p}=0.232$ and $X=0.768$ (\cf Pagel
\& Kazlauskas 1992).  Initial abundances for elements heavier than
helium are initially set to zero.


\begin{table*}[tbh]
\caption[]{ Parameters related to IMF and stellar enrichment}
\begin{tabular}{lll}
\hline
$\gamma$    & $-$2.35      & slope of power-law IMF \\
$(m_{\rm l}$, $m_{\rm u})$ & (0.1, 60) \ms  & stellar mass range at birth \\
$(m_{\rm l}^{\rm AGB}$, $m_{\rm u}^{\rm AGB})$ & (0.8, 8) \ms & progenitor 
mass range for AGB stars \\ 
$(m_{\rm l}^{\rm SNII}$, $m_{\rm u}^{\rm SNII})$ & (8, 30) \ms & progenitor 
mass range for SNII \\
$(m_{\rm l}^{\rm SNIa}$, $m_{\rm u}^{\rm SNIa})$ & (2.5, 8) \ms & progenitor 
mass range for SNIa \\
$\nu^{\rm SNIa}$       & 0.015 & fraction of progenitors ending as SNIa
\\ \hline
\end{tabular}
\end{table*}

We list the main input parameters in Table~2, \ie the adopted
IMF-slope, minimum and maximum stellar mass limits at birth as well as
the progenitor mass ranges for stars ending their lives as AGB star,
SNIa, and SNII, respectively.  For simplicity, we assume the stellar
yields of SNIb,c to be similar to those of SNII. Furthermore, we
assume a fraction $\nu^{\rm SNIa} = 0.015$ of all white dwarf
progenitors with initial masses between $\sim 2.5$ and 8 \ms\ to end
as SNIa.  These and other particular choices for the enrichment by
massive stars are based on similar models recently applied to the
chemical evolution of the Galactic disk (\eg Groenewegen \etal 1995; vdH97).  

\subsection{Spectro-photometric evolution model}

The total luminosity of a galaxy in a wavelength interval
$\Delta \lambda$ is determined by: 1) the stellar contribution 
$L_{\rm *}$, 2) the ISM contribution $L_{\rm ism}$ (\eg H{\sc ii}-regions, high-energy
stellar outflow phenomena, etc.), and 3) the contribution due to absorption and scattering $L_{\rm ext}$:
\begin{equation}
L_{\rm gal}^{\Delta \lambda}(t) = L_{\rm *} + L_{\rm ism} - L_{\rm ext}
\end{equation}
where each term in general is a complex function of galactic evolution
time.  We concentrate on the stellar contribution and neglect the last
two terms in Eq. (1). Then, the luminosity in a wavelength interval
$\Delta \lambda$ at galactic evolution time $t=T$ can be written as:
\begin{eqnarray}
L_{\rm gal}^{\Delta \lambda}(t={T}) = & \\  
\int_{t=0}^{T}  \int_{m_{\rm l}}^{m_{\rm o}({T}-t)}   &
L_{\rm *}^{\Delta \lambda}(m, Z(t), {T}-t)  S(t) M(m)  {\rm d}m 
{\rm d}t \nonumber
\end{eqnarray}
where $m_{\rm l}$ denotes the lower stellar mass limit at birth, 
$m_{\rm o}(t)$ the turnoff mass for stars evolving to their remnant stage
at evolution time $t_{\rm ev}$, and $L_{\rm *}^{\Delta \lambda}$ the luminosity of a 
star with initial mass $m$, initial metallicity $Z(t)$, and age ($T-t$).
We assume a separable SFR: $S(m,t) = S(t)M(m)$ where $S(t)$
is the star formation rate by number [yr$^{-1}$] and $M(m)$ the IMF 
[\ms$^{-1}$]. By convention, we normalise the IMF as 
$\int M(m) {\rm d}m = 1$ where the 
integration is over the entire stellar mass 
range [$m_{\rm l}$, $m_{\rm u}$] at birth (\cf Table~2).

Starting from the chemical evolution model described above, we compute
the star formation history $S(m,t)$, gas-to-total mass-ratio $\mu(t)$,
and age-metallicity relations (AMR) $Z_{i}(t)$ for different elements
$i$.  Thus, at each galactic evolution time $t_{\rm ev}$ the ages and
metallicities of previously formed stellar generations are known.  To
derive the stellar passband luminosity $L_{\rm *}^{\Delta \lambda}$ we
use a set of theoretical stellar isochrones, as well as a library of
spectro-photometric data.  Stellar evolution tracks provide the stellar
bolometric luminosity $L_{\rm *}^{\rm bol}$, effective temperature
$T_{\rm eff}$, and gravity $g$, as a function of stellar age for stars
with initial mass $m$ born with metallicity $Z_{\rm *}$.  We compute
Eq. (2) using a spectro-photometric library containing the stellar
passband luminosities $L_{\rm *}^{\Delta \lambda}$ tabulated as a
function of $T_{\rm eff}$, $\log g$, and $Z_{\rm *}$ (see below).

The turnoff mass $m_{\rm o}({T}-t)$ occuring in
Eq.\ (2) depends on the metallicity $Z(t)$ of stars formed at galactic
evolution time $t_{\rm ev}$. For instance, the turnoff mass for stars born with
metallicity $Z = 10^{-3}$ at a galactic age of $t_{\rm ev} = 14$ Gyr
is $m_{\rm o} \sim 0.8$ \ms\ (\eg Schaller \etal 1992). This value
differs considerably from $m_{\rm o} \sim 0.95$ \ms\ for stars born
with metallicity $Z = Z_{\odot}$. Such differences in $m_{\rm o}$
affect the detailed spectro-photometric evolution of a galaxy by
constraining the mass-range of stars in a given evolutionary phase
(\eg horizontal branch) at a given galactic evolution time.  In the
models described below, we explicitly take into account the dependence
of $m_{\rm o}(t)$ on the initial stellar metallicity $Z_{*}$ (see vdH97)


\begin{figure*}[tbh] 
{\psfig{figure=\lsbfil{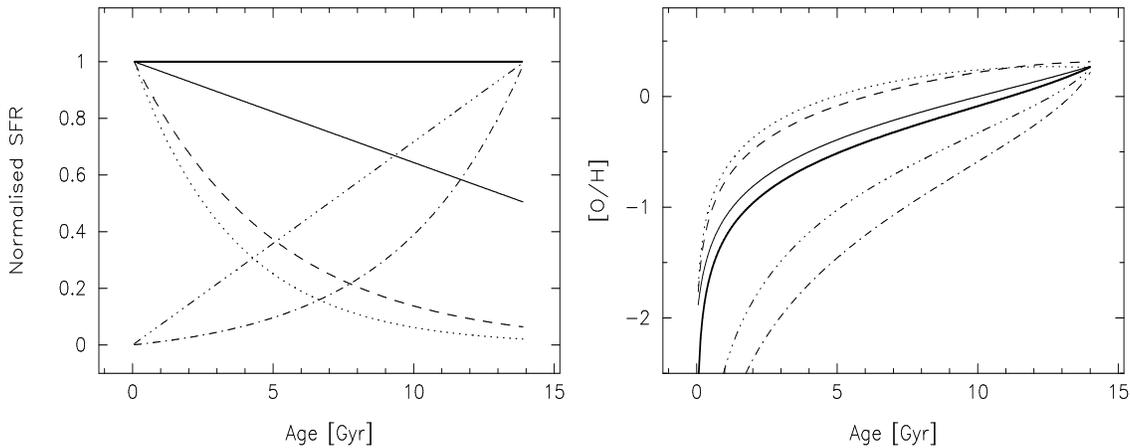},height=7.cm,width=16.cm,angle=270.}}
\caption[]{Theoretical star formation histories ({\bf left}) 
and resulting [O/H] {\it vs} age relations ({\bf right}): constant SFR
(thick solid line), linear decrease (thin solid),
exponential decrease (dashed; $\tau_{\rm sfr} = 5$ Gyr), fast
exponential decrease (dotted; $\tau_{\rm sfr} = 4$ Gyr), linear
increase (dash-dotted), exponential increase
(dot-dashed). SFRs have been normalised to a current gas-to-total
mass-ratio $\mu_{1}=0.1$.}
\end{figure*}


\begin{figure*}
\vbox{
{\psfig{figure=\lsbfil{lsbnumex.ps},height=5.cm,width=14.cm,angle=270.}}
\caption[]{Exponentially decreasing SFR model: total number of stars 
formed ({\em left}) and average stellar bolometric luminosity ({\em
right}) {\it vs} galactic age for different evolutionary phases: MS (solid
curve), RGB (dashed), HB (dot-dashed), EAGB (dotted), and AGB
(dash-dotted).}
\vspace{1.5cm}
\vbox{
{\psfig{figure=\lsbfil{lsbexphdec.ps},height=6.cm,width=14.cm,angle=270.}}
\vspace{0.5cm}
{\psfig{figure=\lsbfil{lsbcnphdec.ps},height=6.cm,width=14.cm,angle=270.}} }
\vspace{0.2cm} 
\caption[]{Photometric evolution of exponentially decreasing SFR
({\em top panels}) and constant SFR ({\em bottom}) models. {\em Left
panels:} $B$-mag contribution for stars in different evolutionary
phases as in Fig.\ 3: MS (solid curve), RGB (dashed), HB (dot-dashed),
EAGB (dotted), AGB (dash-dotted), and Total (thick solid). {\em Right
panels:} Same as left panels but for $I$-mag} }
\end{figure*}

\subsection{Stellar evolution tracks and spectro-photometric data}

We use the theoretical stellar evolution tracks from the Geneva group
(\eg Schaller \etal 1992; Schaerer \etal 1993). These uniform grids
cover large ranges in initial stellar mass and metallicity, \ie $m=
0.05 - 120$ \ms\ and $Z= 0.04 - 0.001$, respectively.  These tracks
imply a revised solar metallicity of $Z_{\odot}=0.0188$ with
$Y_{\odot}=0.299$, and $\Delta Y / \Delta Z = 3.0$ for a primordial
He-abundance of $Y_{\rm p} = 0.232$.  For stars with $m> 7$ \ms ,
these tracks were computed until the end of central C-burning, for
stars with $m= 2-5$ \ms\ up to the early-AGB, and for $m<1.7$ \ms\ up
to the He-flash.  For stars with $m
\la 0.8$ \ms , we used the stellar isochrone program from the Geneva
group (Maeder \& Meynet, priv.\ comm.; $\bull $\bull $\bull$
see also vdH97 $\bull \bull \bull$).

To cover the latest stellar evolutionary phases (\ie horizontal branch
(HB), early-AGB, and AGB) for stars with $m \la 8$ \mss, we extended
the tracks from Schaller \etal with those from Lattanzio (1991; HB and
early-AGB) for $m \sim 1-2$ \mss, and from Groenewegen \& de Jong
(1993; early-AGB and AGB) for $m \sim 1-8$ \mss. These tracks roughly
cover the same metallicity range as the tracks from the Geneva group.
Corresponding isochrones were computed on a logarithmic grid of stellar
ages, covering galactic evolution times up to $t_{\rm ev} \sim 14$
Gyr.  Isochrones are linearly interpolated in $m$, log $Z$, and log
$t$.

The spectro-photometric data library that we use is based on the
Revised Yale Isochrones and has been described extensively by Green
\etal (1987). These data include stellar $UBVRI$ Johnson-Cousins
magnitudes covering the following ranges in $T_{\rm eff}[{\rm K}] =
2800$ to 20000, log $g$ [cm s$^{-2}$] $= -0.5$ to 6, and log ($Z/
{Z}_{\odot}) \sim -2.5$ to $+0.5$. Corresponding spectro-photometric
data for stars with $T_{\rm eff} > 20000$ K have been adopted from
Kurucz (1979) at solar metallicity, covering $T_{\rm eff} = 20000 - 50000$ K.

Although a detailed description of the tuning and calibration of the
adopted photometric model is beyond the scope of this paper, we note
that the model has been checked against various observations including
integrated colours and magnitudes, luminosity functions, and
colour-magnitude diagrams of Galactic and Magellanic Cloud open and
globular clusters covering a wide range in age and metallicity
(see vdH97).

\section{General properties of the model}

Before discussing the modelling of LSB galaxies, we first describe the
general behaviour of the models assuming a few different star formation
scenarios.

We start from a model galaxy with initial mass $M_{\rm g}(t=0) =
10^{10}$ \mss, initially metal-free and devoid of stars.  The chemical
and photometric evolution of this galaxy is followed during an evolution
time $t_{\rm ev} = 14$ Gyr, assuming one of the theoretical star
formation histories discussed below. Unless stated otherwise, we
assume that stars are formed according to a Salpeter (1955) IMF (\ie
$\gamma = -2.35$) with stellar mass limits at birth between 0.1 and 60
\ms\ (\cf Table~2).


\begin{table*}[bth]
\caption[]{Star formation models ($\mu_{1} = 0.1$ and $M_{\rm 
tot} = 10^{10}$ \mss, unless noted otherwise)}
\begin{tabular}{llllllcclll}
\hline
(1) & (2) & (3) & (4) & (5) & (6) & (7) & (8) & (9) & (10) & (11)\\
 & Model & $\langle$SFR$\rangle$ &  SFR$_{1}$ & $\alpha$ & [O/H]$_{1}$ & $N_{\rm tot}$ & 
$L_{B}$ & $M_{\rm HI}/L_{B}$ & $\gamma$ & Notes\\
& & \multicolumn{2}{c}{[\ms yr$^{-1}$]} & & & $[\times 10^{10}]$& [$\times 10^9 L_{\odot ,{B}}$] & 
[$M_{\odot}/L_{\odot , {B}}$] & &\\ \hline
A1 & Exp. decreasing 
&  0.93 & 0.17 & 0.18 & +0.3 & 3.6 & 3.6  & 0.16 & 
$-2.35$ & (a) \\
A2 & "          &  0.52 & 0.09 & 0.18 & $-$0.25 & 2.1 & 2.3  & 1.56 & 
$ -2.35$ &(a, b) \\
A3 & "          &  0.69 & 0.13 & 0.18 & $-$0.9 & 4.7 & 1.6  & 0.45 & 
$ -3$  &(a) \\
& & & & & & & & & & \\
B1 & Constant   &  0.89 & 0.89 & 1.0   & +0.3 & 3.4 & 10 & 0.03 & 
$-2.35$ &\\
B2 & "          &  0.49 & 0.49 & 1.0   & $-$0.25 & 2.1 & 6.4  & 0.58 & 
$ -2.35$& (b) \\
B3 & "          &  0.68 & 0.68 & 1.0   & $-$0.9 & 4.7 & 3.7  & 0.21 & 
$ -3$& \\
& && & & & & & & & \\
C  & Exp. decreasing  &  0.90 & 0.07 & 0.08 & +0.3 & 11 & 2.3  
& 0.32 & $-2.35$ & (c) \\
D  & Lin. decreasing &  0.90 & 0.60 & 0.67 & +0.3 & 3.5 & 7.6 & 0.10 
& $-2.35$ &\\
E  & Exp. increasing &  0.85 & 3.06 & 3.4  & +0.3 & 3.3 & 28 & 0.03 
& $-2.35$ &\\
F  & Lin. increasing &  0.87 & 1.74 & 2.0  & +0.3 & 3.3 & 18 & 0.04 
& $-2.35$ &\\ 
\hline 
\end{tabular}

\vskip 2pt
{\bf Notes:} (a) $\tau_{\rm sfr} = 5$ Gyr. (b) $\mu_{1} = 0.5$. (c) $\tau_{\rm sfr} = 4$ Gyr\\
\end{table*}

The following star formation histories are considered: 1) constant, 2)
exponentially decreasing, 3) linearly decreasing, 4) exponentially
increasing, and 5) linearly increasing.  Normalised SFRs and resulting
age-ISM metallicity relations are shown in Fig.\ 2.  For each model, the
amplitude of the SFR is chosen such that a present-day gas-to-total
mass-ratio $\mu_{1} = 0.1$ is achieved.

In columns (2) to (6) of Table~3, we list the functional form of the
SFR, average past and current SFRs, the ratio of current and average
past SFRs $\alpha$, and the present-day mass averaged ISM oxygen
abundance.  $\bull $ The average past SFR is roughly the same for all
models ending at $\mu_1 = 0.1$ and is $\langle$SFR$\rangle = 0.9$ \ms
yr$^{-1}$.  In contrast, {\em present-day} SFRs range from SFR$_{1}$ =
0.07 to 3 \ms yr$^{-1}$ and in fact determine the contribution by
young stars to the integrated light of the model galaxy. Current
oxygen abundances predicted are [O/H]$_{1} \sim +0.25$ and are mainly
determined by $\mu_{1}$, the IMF, and the assumed mass limits for SNII
(\cf Table~3).  Column (7) gives the number of stars formed, Column
(8) their total $B$ band luminosity. Column (10) shows the resulting
H{\sc i} mass-to-light ratio. Column (11) gives the slope of the IMF
used, while Column (11) refers to additional Notes.

Let us consider the photometric evolution of the constant and
exponentially decreasing SFR models in some more detail.  Fig.\ 3
shows the evolution of the total number of MS and post-MS stars for
exponentially decreasing SFR model A1. The current total number of MS
stars is roughly $4\cdot 10^{10}$, compared to $\sim 10 ^{8}$ post-MS
stars (all phases) and $\sim 3\cdot 10^{4}$ for AGB stars only.  The
present-day mean luminosity of stars on the MS is $\sim 10^{-1}\
L_{\odot}$ compared to $\sim 10^{4}\ L_{\odot}$ for AGB stars.  The
current bolometric galaxy luminosity is determined mainly by MS stars
($L_{\rm MS} \sim 4 \cdot 10^{9}\ L_{\odot}$). In particular, AGB
stars ($L_{\rm AGB} \sim 10^{8}\ L_{\odot}$) are relatively
unimportant.


\begin{figure}
{\psfig{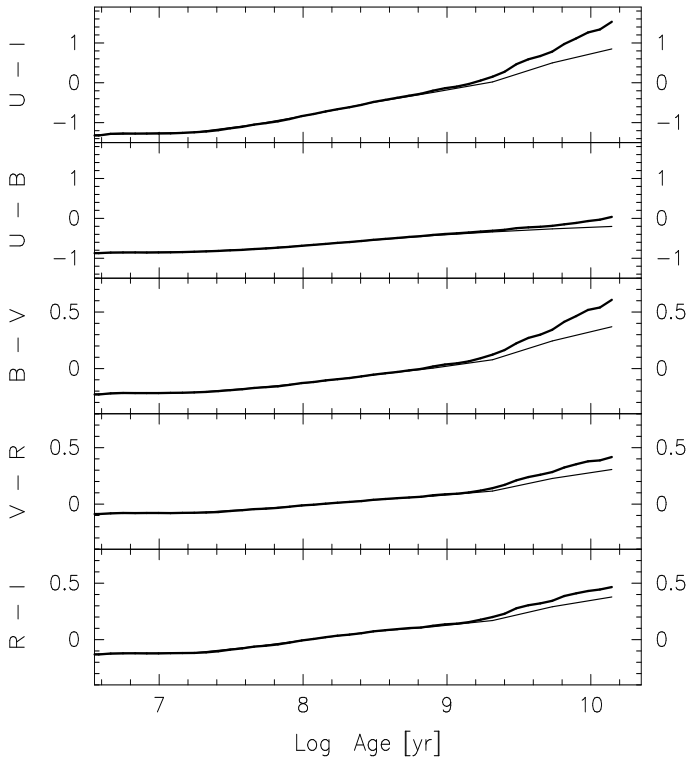}}
\caption{Broadband colours {\it vs} galactic age for exponentially
decreasing SFR (thick solid) and constant SFR (solid) models. Note the
different magnitude scales in the upper two and bottom three panels,
respectively.}
\end{figure}
We show in Fig.\ 4 the $B$ and $I$-band magnitudes of stars in
different evolutionary phases (again for the exponential model).  As
for the bolometric galaxy luminosity, MS stars generally dominate in
the $B$-band.  However, in the $I$-band, RGB and HB stars are nearly
as important as MS stars, at least in later stages of galactic
evolution. Due to the cooling of old, low-mass MS stars as well as the
increasing contribution by RGB and HB stars with age, the current
total $I$-band magnitude is considerably brighter than that in the
B-band. This qualitative model behaviour is insensitive to the adopted
star formation history (for {\em e}-folding times larger than $3-5$
Gyr) but instead is determined by the assumed IMF and the stellar
input data used. 
Thus, constant star
formation models exhibit a similar behaviour apart from being brighter
by about one magnitude in all passbands at late times (\cf Fig.\ 4).

Fig.\ 5 illustrates the sensitivity of broadband colours to the
star formation history for constant and exponentially
decreasing SFR models. The colours considered here redden  with
galactic age. In general, differences between colours such as for
different SFR models are less than the variations of these colours with
age for a given model.  Assuming a galactic age of \eg 8 instead of 14
Gyr has limited effect on the resulting galaxy colours (\eg less than
0.1 mag in $B-V$), even though absolute magnitudes are substantially
altered.  Both age and extinction effects can result in
substantial reddening of the colours of a stellar population in almost
the same manner and it is difficult to disentangle their effects on
the basis of photometry data alone.  Clearly, galaxy colours {\em
alone} are not well suited to discriminate between different SFR
models, even when internal extinction is low and other reddening
effects are negligible (see below).

\section{Results}

As our emphasis will be on a comparison between the evolution of HSB
and LSB galaxies, we will first focus on the relatively well-known
evolution of HSB galaxies, using a model chosen to emphasise the
observed properties of HSB galaxies.  Using this as a starting point,
we explore other models to find the best model for LSB galaxies.

\subsection{HSB galaxies: $\mu_1 = 0.1$ models}

Models using exponentially decreasing SFRs and different {\em
e}-folding times for different Hubble types (\eg Larson
\& Tinsley 1978; Guiderdoni \& Rocca-Volmerange 1987; Kennicutt 1989;
Bruzual \& Charlot 1993; Fritze-v. Alvensleben \& Gerhard 1994) are
usually found to be in good agreement with observations of HSB
galaxies.  Therefore we will mainly focus on the exponential SFR
models discussed above.  As the gas-richness is one of the obvious
differences between HSB and LSB galaxies, we will first focus on
models with $\mu_1 = 0.1$ to describe HSB galaxies.

Fig.\ 6 shows the results of the exponentially decreasing SFR model:
SFR $\propto \exp (-t/ \tau_{\rm sfr})$ with $\tau_{\rm sfr} = 5$ Gyr
ending at $\mu_{1} = 0.1$ at $t=t_{ev}=14$ Gyr for several initial
galaxy masses $M_{\rm g}(t=0) \sim 10^{8}-10^{10}$ \mss .
%
The precise value of $\mu_{1}$ was achieved by scaling the
amplitude of the SFR accordingly.  Galaxy colours and abundances are
not affected by this scaling while absolute magnitudes and final gas
masses scale with $M_{\rm g}(t=0)$ (as indicated in Fig.\ 6).

Present-day ($B-V$) and ($R-I$) colours for this model are 0.6 and 0.5,
respectively.  HSB galaxies with $(B-V)\ga 0.6$ and $(R-I)\ga 0.5$ can
be explained only when substantial amounts of internal dust extinction
are incorporated.  The reason is that even though a single stellar
population born with metallicity $Z=0.02$ may become as red as $(B-V)
\sim 1.1$ and $(R-I) \sim 0.8$, the luminosity contribution by young
stellar populations (with ages less than a few Gyr) results in
substantial blueing of the galaxy colours. For the same reason, values
of $(B-V) \ga 0.65$ are not predicted by dust-free models independent
of the adopted (exponential) SFR (as long as $\tau \ga 4$ Gyr) or
IMF. Therefore, considerable reddening of the {\em younger} stellar
populations is required to explain the colours of HSB galaxies in this
model.  Figs.~6c and d show that both the [O/H] abundance ratios and
present-day gas masses of HSB galaxies can be explained reasonably
well for $M_{\rm g}(t=0) > 10^{10}$ \mss.  These conclusions agree
with results from many previous models (with different stellar input
data) for HSB galaxies (\eg Larson \& Tinsley 1978; Guiderdoni \&
Rocca-Volmerange 1987).

It is clear from Fig.\ 6 that this model is not able to explain the
properties of the LSB galaxies in our sample. The model colours are too
red while the predicted [O/H] abundances are too high. As LSB galaxies
have higher gas-fractions we thus need to broaden the range of
gas-fractions.  

\subsection{LSB galaxies: $\mu_1 \neq 0.1$ models}

\begin{figure*}[t!]
{\psfig{figure=\lsbfil{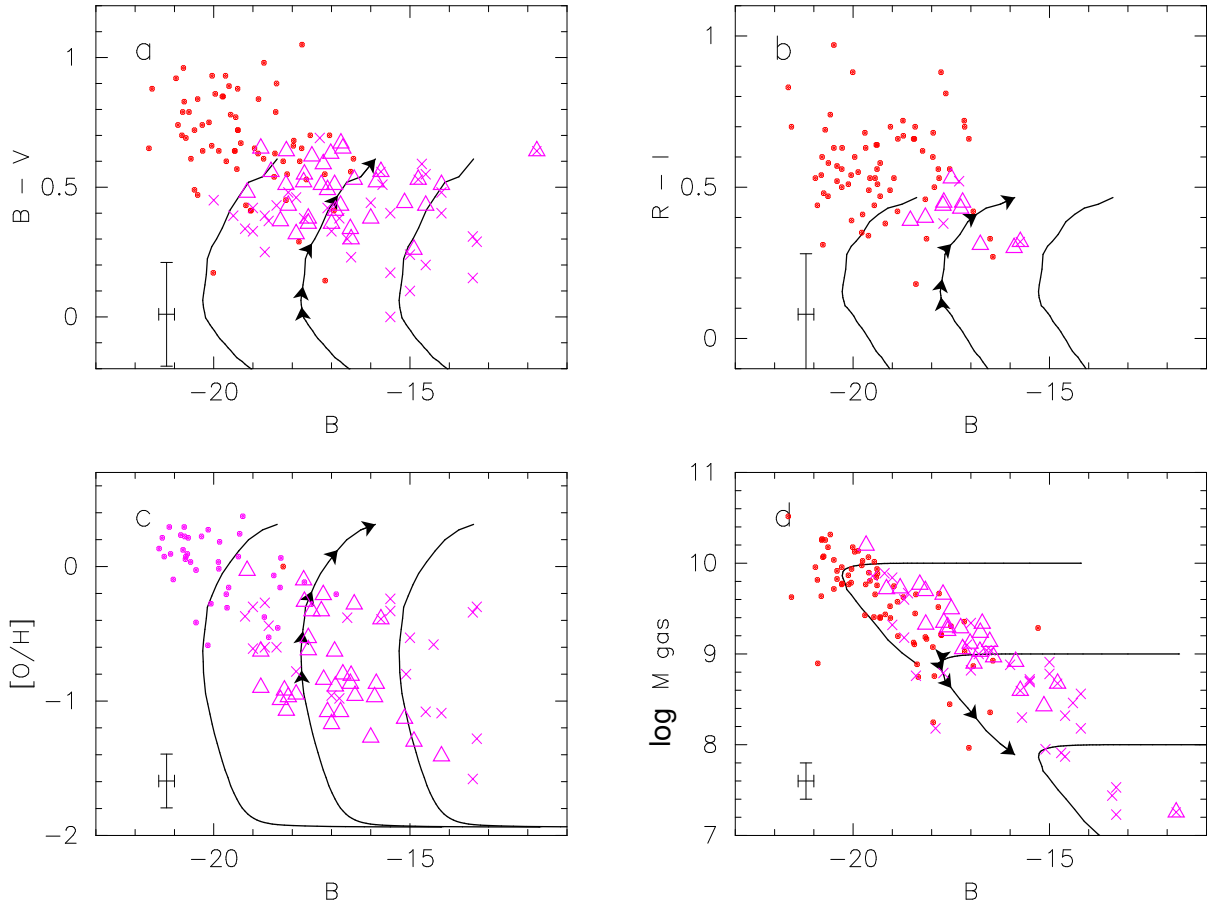},height=12.cm,width=16.cm,angle=270.}}
\caption[]{Exponentially decaying SFR model A1: photometric and chemical 
evolution results. Curves have been drawn for initial galaxy masses
$M_{\rm g}(t=0) = 10^{8}$, $10^{9}$, and $10^{10}$ \mss.  For the
$M_{\rm g}(t=0)= 10^{9}$ \ms\ model, arrows indicate evolution times
of 1, 2, 4, 8, and 14 Gyr, respectively. Symbols refer to the
following galaxy samples: face-on HSB spirals (dots; de Jong \& van
der Kruit 1994), LSB spiral galaxies (triangles; de Blok \etal 1995),
and dwarf irregulars (crosses; Melisse and Israel 1994). Typical
observational errors are shown in the bottom left of each panel.}
\end{figure*}


\begin{figure*}
{\psfig{figure=\lsbfil{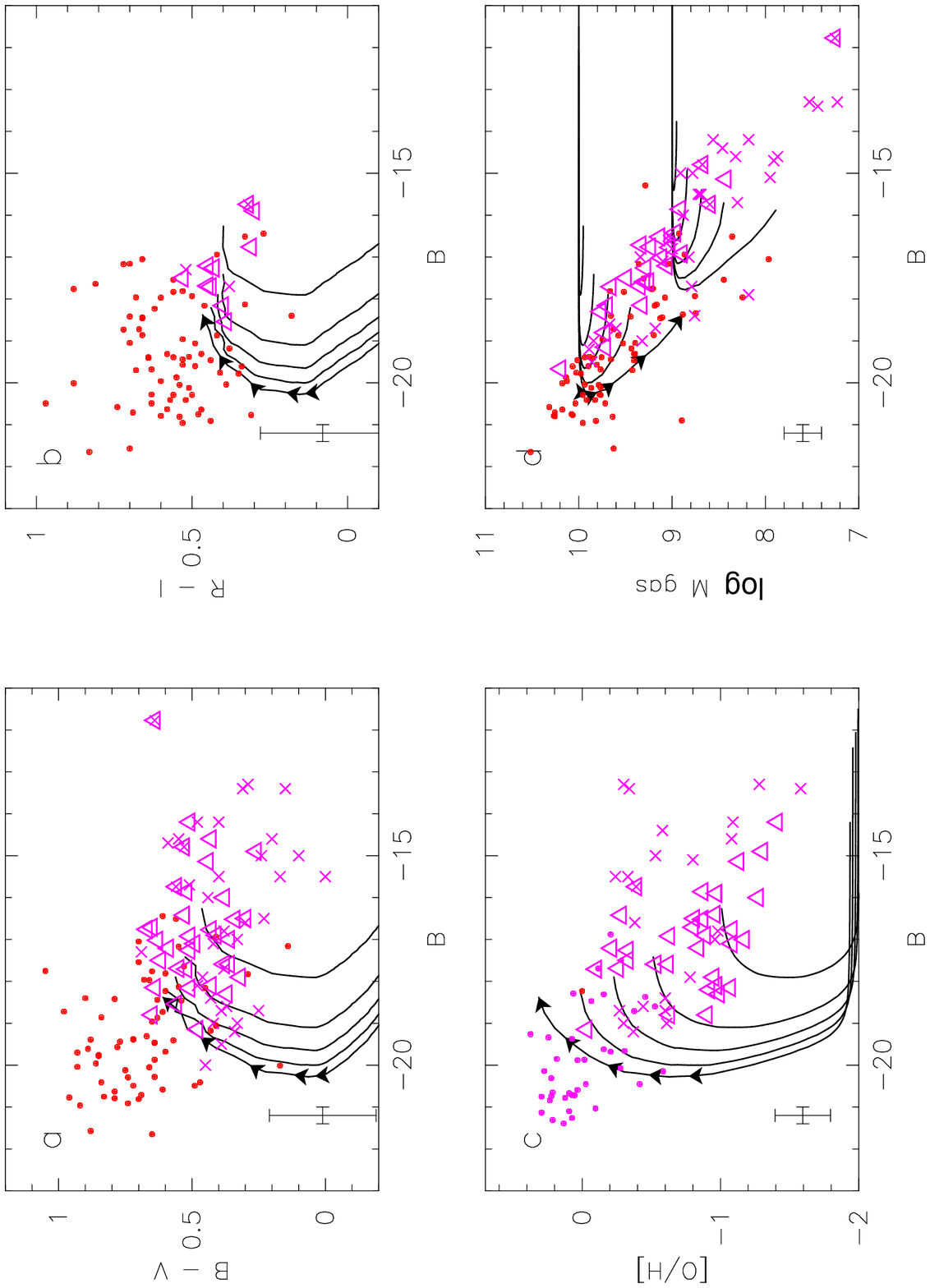},height=12.cm,width=16.cm,angle=270.}}
\caption{Photometric and chemical evolution results in case of 
exponentially decaying SFR models ending at $\mu_{1} =$ 0.025, 0.3,
0.5, 0.7, and 0.9 for an initial galaxy mass of $M_{\rm g}(t=0)=
10^{10}$ \ms\ (in Fig.\~7d we show also results for $M_{\rm g}(t=0)=
10^{9}$ \mss).  Arrows on top of the $\mu_{1} = 0.025$ model indicate
evolution times of 1, 2, 4, 8, and 14 Gyr, respectively. Observational
data as in Fig.~6}
\end{figure*}

We now consider models with gas fractions $\mu_{1} =$ 0.025, 0.1, 0.3,
0.5, 0.7, and 0.9. First we will examine the constraints which {\it
individual} properties such as abundance and gas-fraction impose on
the models, before combining these in a final model.

\subsubsection{Colours and magnitudes}

Results for a model with initial mass $M_{\rm g}(t=0)= 10^{10}$ \ms\
are shown in Fig.~7.  While the age distribution of the stellar
populations is identical in all the models shown, present-day ($B-V$) and
($R-I$) colours are found to decrease by 0.2 and 0.1 mag, respectively,
when going from models ending at $\mu_{1} = 0.025$ to $0.9$. This
bluing effect is due to the decrease of stellar metallicities for
models ending at increasingly higher gas fractions.

The LSB galaxies can be distinguished in two major groups by means of
their colours. First, LSB galaxies with $(B-V) \ga 0.5$ which can best
be modelled with exponentially decaying SFR models ending at values
$\mu_{1} = 0.5 - 0.7$ at age 14 Gyr.  These results can be shifted
towards brighter or fainter $B$ magnitudes by changing the initial gas
masses accordingly.  This leaves the resulting colours unaltered (\cf
Fig.~6).

Second, relatively blue LSB galaxies with $(B-V) \la 0.4$ cannot be
fitted by exponentially decreasing SFR models with $\tau \la 5$ {\em
alone} (assuming $t_{\rm ev} = 14$ Gyr), regardless of their current
gas fraction $\mu_{1}$.  There are several possible explanations: a
relatively young ``surplus'' stellar population may influence the
galaxy colours resulting from an underlying exponentially decreasing
SFR. Alternatively, these LSB galaxies may have started forming stars
recently (\eg $5-8$ Gyr instead of 14 Gyr, \cf Fig.~7a), or they may
have more slowly decaying or constant SFRs which generally result in
present-day colours of $(B-V) \la 0.4$ and $(R-I) \la 0.35$ (but the
latter option appears to be excluded because of their abundances and
gas fractions; see below).

\subsubsection{Abundances}

The observed range in [O/H] abundances for LSB galaxies is well
explained by exponentially decaying SFR models with $\mu_{1} \ga 0.3$
(\cf Fig.~7c).  Constant SFR models ending at $\mu_{1}
\ga 0.3$ are also consistent as the abundances of elements
predominantly produced in massive stars are in general determined by
the present-day gas fraction $\mu_{1}$ and are insensitive to the
detailed underlying star formation history (\eg Tinsley 1980).
However, these latter models appear to be excluded by the measured gas
masses (see 5.2.4 and 5.2.5) { (for a closed box model)}.  Though the
abundances of metal-poor LSB galaxies with [O/H] $\la -1$ can be well
explained by models ending at $\mu_1 =0.9$, it is likely that the
exponential models presented here are an over-simplification for these
gas-dominated systems. It is more likely that they have experienced a
low and sporadic star formation rate.

\subsubsection{Gas content}

Present-day gas masses observed in LSB galaxies can be well fitted by
exponentially decaying SFR models ending at $\mu_{1} \sim 0.5$ (Fig.\
7d).  This is consistent with the range of $\mu_{1} \ga 0.3$ derived
from the [O/H] data. Exponentially decreasing SFR models ending at
$\mu_{1} \ga 0.6$ are inconsistent with the observations unless we
assume that LSB galaxies have started forming stars only a few Gyr
ago.

Fig.~8 displays the gas mass {\it vs} $(B-V)$ for constant and
exponentially decaying SFR models.  Exponentially decreasing SFR
models are able to explain simultaneously values of $\mu_{1} \sim 0.5
\pm 0.2$, $(B-V) \ga 0.5$, and $M_{\rm g}
\ga 10^{9}$ \mss, as observed for the majority of the LSB galaxies in
our sample. Over the entire range of possible gas fractions, constant
(or increasing) SFR models are { only able to explain the bluest galaxies.}

\begin{figure}[b!]
\vbox{ 
\psfig{figure=\lsbfil{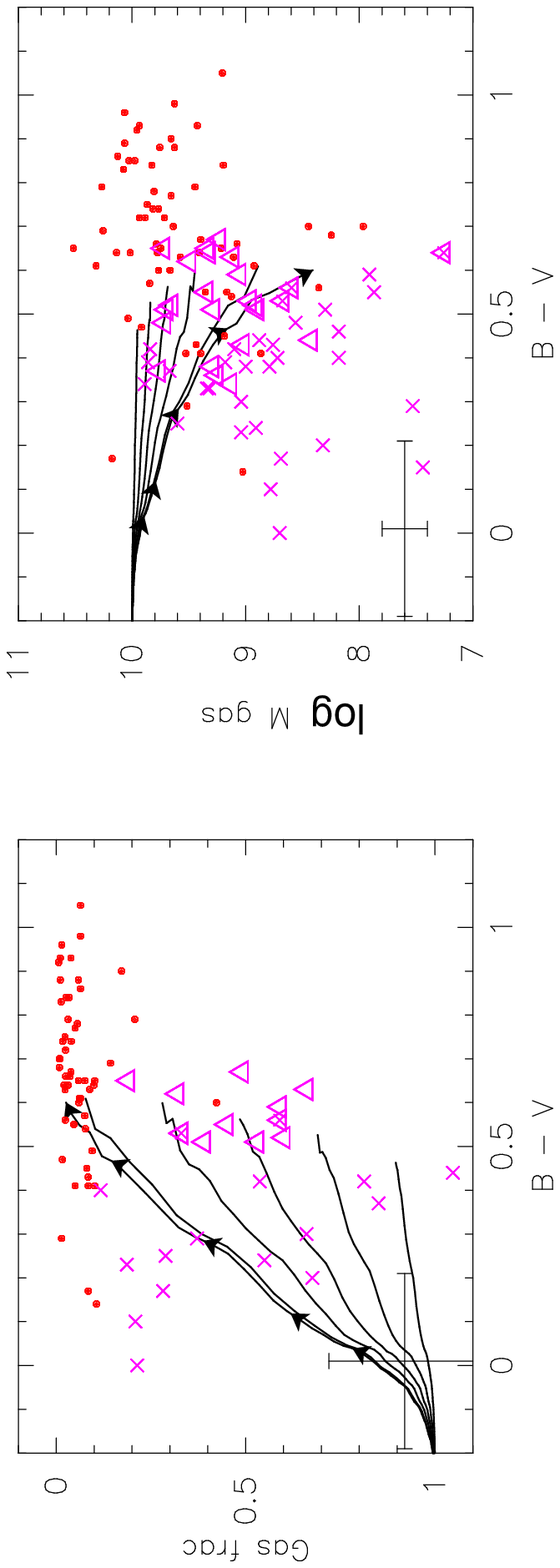},height=7.cm,width=8.cm,clip=t,angle=270.}
\psfig{figure=\lsbfil{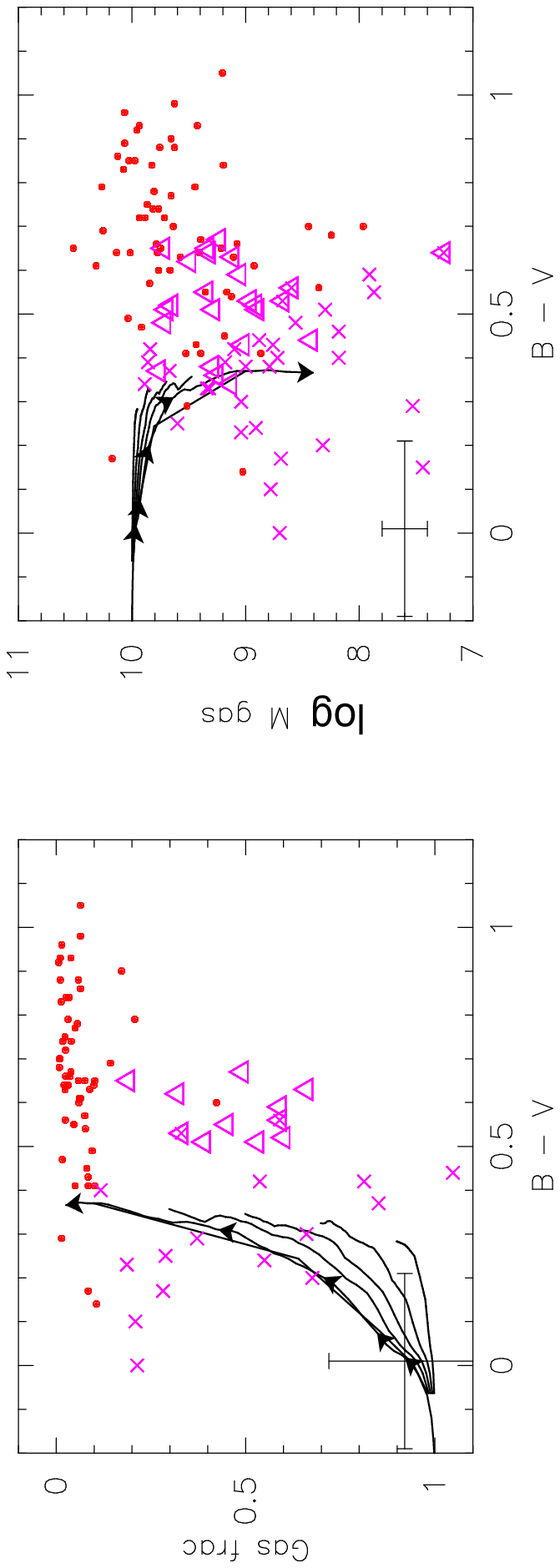},height=7.cm,width=8.cm,clip=t,angle=270.}
}
\caption[]{Comparison of predicted and observed 
gas masses {\it vs} $(B-V)$. {\em Top:}
exponentially decreasing SFR models. {\em Bottom:} constant SFR
models.  Models with SFR normalisations according to $\mu_{1} =$
0.025, 0.1, 0.3, 0.5, 0.7, and 0.9 are shown for an initial galaxy
mass of $M_{\rm g}(t=0)= 10^{10}$ \mss. Arrows on top of the $\mu_{1}
= 0.025$ model indicate evolution times of 1, 2, 4, 8, and 14 Gyr,
respectively. Observational data as in Fig.~6}
\end{figure}

Fig.~8 demonstrates that exponentially decreasing SFR models ending at
$\mu_{1} \sim 0.5$ are in good agreement with the observed hydrogen
mass-to-light ratios $M_{{\rm HI}}/ L_{B}$, in contrast to constant
(or slowly decreasing) SFR models (expect for the bluest systems).

The conclusion that constant and slowly decreasing SFR models are
inconsistent with the observed photometric and chemical properties of
LSB galaxies is based on the assumption of negligible amounts of dust
extinction in these systems. However, if a considerable fraction of
the LSB spirals in our sample would suffer from extinctions of
$E_{(B-V)} = 0.1-0.25$, the $M_{{\rm HI}}/L_{B}$ ratios predicted by
constant SFR models would increase by a factor $1.4-2.5$ after
correction for internal extinction. In this manner, constant and
slowly decreasing SFR models ending at gas fractions $\mu_{1} =
0.3-0.7$ after $\sim 14$ Gyr could also explain relatively red LSB
galaxies with $(B-V) \ga 0.5$ and $M_{{\rm HI}}/L_{B}$ ratios as large
as $\sim 1.5$. Even though we argued in Sect.~2 that internal
extinction in LSB galaxies is unlikely to exceed $E_{(B-V)}
\sim 0.1$, knowing the dust content of LSB galaxies is obviously of crucial
importance in modelling their evolution.

\subsubsection{Summary: the best model}

To summarise, we find that exponentially decreasing SFR models ending
at $\mu_{1} = 0.3-0.7$ are in agreement with the colours, magnitudes,
[O/H] abundances, gas contents, and mass-to-light ratios observed for
LSB galaxies with $(B-V) \ga 0.45$.  Blue LSB galaxies with $(B-V) \la
0.45$ cannot be fitted by exponentially decreasing SFR models without
an additional light contribution from a young stellar
population. Alternatively, such LSB galaxies may have experienced
constant SFRs, or may be much younger than 14 Gyr.

\subsection{HSB/LSB: Just an extinction effect?}
 
Since the models require an internal extinction in HSB galaxies up to
$E_{(B-V)} \sim 0.5$ (corresponding to $A_{V}\sim 1.5$ mag),
independent of the SFH, one could argue that extinction may be the
main explanation for the difference in colour observed between LSB and
HSB galaxies.


As shown in this paper the models indicate that the underlying stellar
populations in LSB and HSB galaxies are still distinctly different due
to differences in the chemical evolution of these systems. Therefore,
extinction effects, although important, cannot be the {\it entire}
explanation for the colour differences observed.

\section{The effects of small star formation bursts}

\subsection{Luminosity contribution by HII regions in LSB 
galaxies}

The presence of H{\sc ii} regions in all the LSB galaxies in our
sample suggests that recent star formation is a common phenomenon in
these systems, and may thus influence the observed properties. In the
previous section we found that for some LSB galaxies a simple
exponential SFR was unable to explain the observed properties and that
an additional ``burst'' of young stars was needed. Here we examine the
effects of such bursts on the colours of LSB galaxies. To this end, we
selected all H{\sc ii} regions that could be identified by eye, either
from H$\alpha$ or $R$-band CCD images, and added their total
luminosity in the $I$-band. We restrict ourselves to the $I$-band data
as to be least affected by extinction that may be present in or near
H{\sc ii} regions (McGaugh 1994).
We define $\eta_{I}$ as the
ratio of this H{\sc ii} region integrated luminosity and the total
luminosity of the corresponding LSB galaxy.  As we are probably
incomplete at low H{\sc ii} region luminosities, this ratio gives a lower
limit for the H{\sc ii} region contribution to the total luminosity.


\begin{table*}[th]
\caption[]{Observed broadband contributions by H{\sc ii} regions in LSB 
galaxies and derived SFRs}
\begin{tabular}{llllcclll}
\hline
(1) & (2) & (3) & (4) & (5) & (6) & (7) & (8) & (9) \\
Name & \# & $I$ & $\eta_{\rm I}$ & 
$(B-V)_{{\rm HII}}$ & $(B-V)_{\rm gal}$ & SFR$^{\rm cont}$ &
SFR$^{\rm burst}$ & SFR$^{\rm tot}$ \\ 
& & [mag]& & [mag] & [mag] & \multicolumn{3}{c}{[M$_{\odot}$ yr$^{-1}$]} 
\\ \hline
F561-1  & 15 & $-$15.3 & 0.06 & 0.43 & 0.59 & 0.013 & 0.068 & 
0.081 \\
F563-1  & 12 & $-$14.1 & 0.12\dag & 0.65 & 0.65 & 0.16 & 1.53 
& 1.7\dag 
\\
F563-V1 & 4  & $-$13.6 & 0.04 & 0.53 & 0.56 & 0.005 & 0.019 & 
0.024 \\
F564-V3 & 16 & $-$10.1* & 0.10* & 0.78 & 0.64 & $-$ & $-$  & $-$ \\
F565-V2 & 4  & $-$12.2* & 0.04* & 0.72 & 0.53 & 0.017 & 0.055 
& 0.072 \\
F567-2  & 9  & $-$14.5 & 0.05\dag & 0.90\dag & 0.67 & 0.029 & 
0.12 & 0.15 \\
F568-1  & 4  & $-$14.7 & 0.02 & 0.41 & 0.62 & 0.12 & 0.19 & 0.31 \\
F568-3  & 7  & $-$16.1 & 0.07 & 0.53 & 0.55 & 0.06 & 0.27 & 0.33 \\
F568-V1 & 11 & $-$16.8 & 0.22 & 0.59 & 0.51 & 0.05 & 0.54 & 0.58 \\
F571-5  & 10 & $-$14.6* & 0.10* & 0.28 & 0.34 & $-$ & $-$ & $-$ \\
F571-V1 & 5  & $-$13.9* & 0.04* & 0.40 & 0.53 & 0.033 & 0.11 & 
0.14 \\
F574-2  & 4  & $-$14.0* & 0.03* & 0.46 & 0.63 & 0.012 & 0.029 
& 0.041 \\
F577-V1 & 14 & $-$16.3* & 0.19* & 0.39 & 0.38 & $-$ & $-$ & $-$ \\
U128    & 17 & $-$16.2 & 0.06 & 0.55 & 0.51 & 0.14 & 0.68 & 0.82 \\
U628    & 26 & $-$17.8 & 0.17 & 0.55 & 0.56 & $-$ & $-$ & $-$ \\ 
U1230   & 24 & $-$16.2 & 0.08 & 0.43 & 0.52 & 0.05 & 0.34 & 0.39   
\\ 
\hline 
\end{tabular}

{\bf *} values refer to $R$ band magnitudes instead  of $I$-band \\
{\bf \dag} uncertain due to contamination by fore- or background objects \\
{\bf ---} $\mu_{\rm rot}$ not determined
\end{table*}

In columns (1) and (2) of Table~4, we list the LSB galaxy
identification and number of H{\sc ii} regions selected. The number of
H{\sc ii} regions identified within individual LSB galaxy ranges from a few
to $\sim 25$.  For the ensemble of H{\sc ii} regions in each LSB
galaxy, we tabulate the absolute $I$ magnitudes as well as the
corresponding ratios  $\eta_{I}$ of the H{\sc ii} region
integrated luminosity and total LSB galaxy luminosity, in columns (3)
and (4). Mean ($B-V$) colours for the H{\sc ii} regions and for the LSB
galaxy as a whole are given in columns (5) and (6), respectively.

For most of the LSB galaxies in our sample, the contribution of the
H{\sc ii} regions to the total light emitted by LSB galaxies does not
exceed $\eta_I = 0.05 - 0.1$. However, as H{\sc ii} regions are likely
to contain an increased amount of dust (McGaugh 1994) the actual
contributions may be higher by a factor of $\sim 2$ in $I$.  Thus, the
values of $\eta_{I}$  provide lower limits to the actual
luminosity contributions of the H{\sc ii} regions.  For some LSB
galaxies, \eg F568-V1 and F577-V1, the H{\sc ii} region contribution
is found to be as high as $\eta_{I} = 0.2$. These systems contain a
only modest number of H{\sc ii} regions so that their H{\sc ii}
regions on average may be larger and/or brighter than those present in
several other LSB galaxies.

Figure~10 shows the resulting H{\sc ii} region contributions
$\eta_{I}$ for the SFR models discussed in Sect.~4. If we assume a
maximum age $\tau_{\rm HII} = 5$ Myr for the H{\sc ii} regions
observed, then this implies that stars more massive than $\sim 25$
\ms\ are associated with these regions, according to the stellar
evolution tracks from the Geneva group (see below).

Corrections for dust extinction within the H{\sc ii} regions and LSB
galaxy as a whole, respectively, will shift the observations in the
directions as indicated in Fig.~10 (assuming a mean Galactic
extinction curve). From the $I$ band observations we can conclude that
the values of $\eta_{I}$ predicted by smoothly varying SFR models are
systematically too low for most of the LSB galaxies in our sample.

This is true in particular for F568-V1, F577-V1, and U628 for which
values of $\eta_{I, R} \ga 0.2 - 0.25$ suggest that star formation has
been recently enhanced by factors $\sim 5-10$ relative to the SFRs
predicted by exponentially decreasing or constant SFR models.

\subsection{Effects of small amplitude star formation bursts}

We investigate whether the observed values of $\eta_{I} \sim 0.25$ in
the $I$ band, as observed for several LSB galaxies discussed above,
can be explained by small-amplitude bursts of star formation.

We assume a Gaussian star formation burst profile with a given
amplitude $A_b$ and dispersion $\sigma$.  We follow its evolution
during 1 Gyr with a time resolution of $\sim 0.1$ Myr at time of burst
maximum and of $\sim 2$ Myr at roughly 5$\sigma$ from burst maximum.
We superimpose the star formation burst on an exponential SFR model,
as discussed in Sect.~5.1.

We initially assume a galactic evolution time at burst maximum of
$t_{\rm b} = 13$ Gyr, burst maximum amplitude $A_{\rm b} = 8$ \ms\,
yr$^{-1}$, burst duration $\Delta t_{\rm b} = 5$ Myr, maximum H{\sc
ii} region lifetime $\tau_{\rm H{II}} = 5$ Myr, and an initial galaxy
mass of $10^{10}$ \mss.  For simplicity, we neglect any influence of
the burst on the chemical evolution of the model galaxy.


\begin{figure*}[thb] 
\vbox{\psfig{figure=\lsbfil{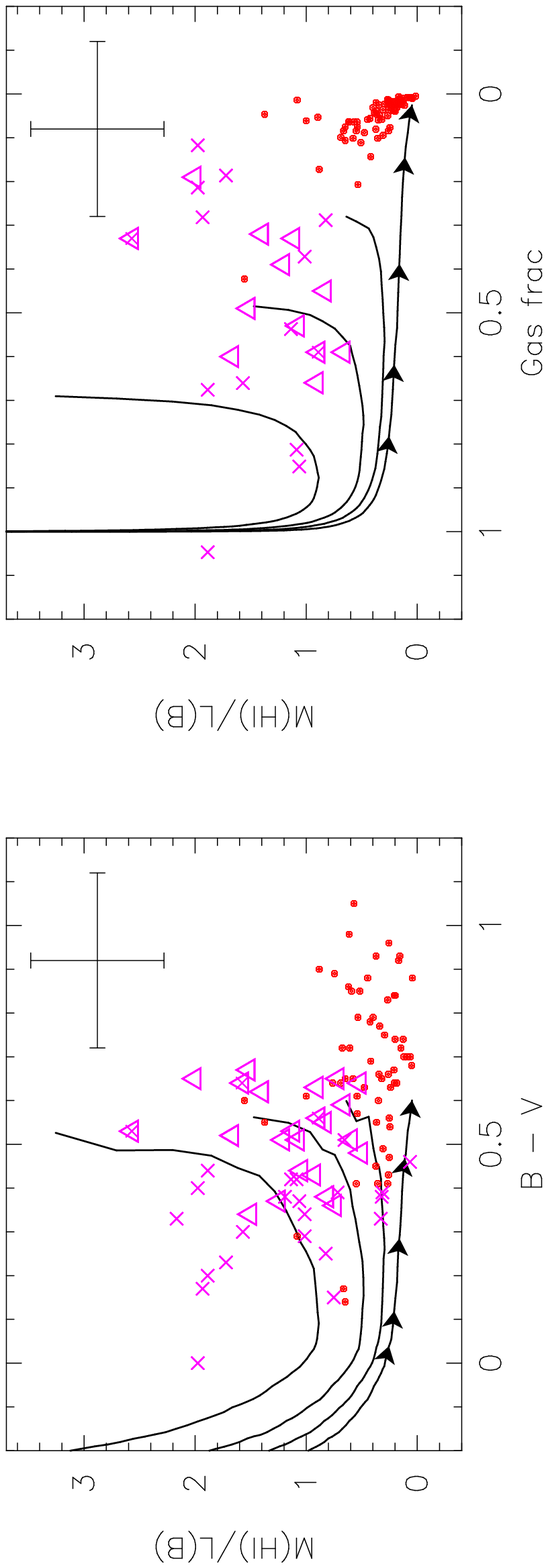},height=7.cm,width=16.cm,angle=270.}
\psfig{figure=\lsbfil{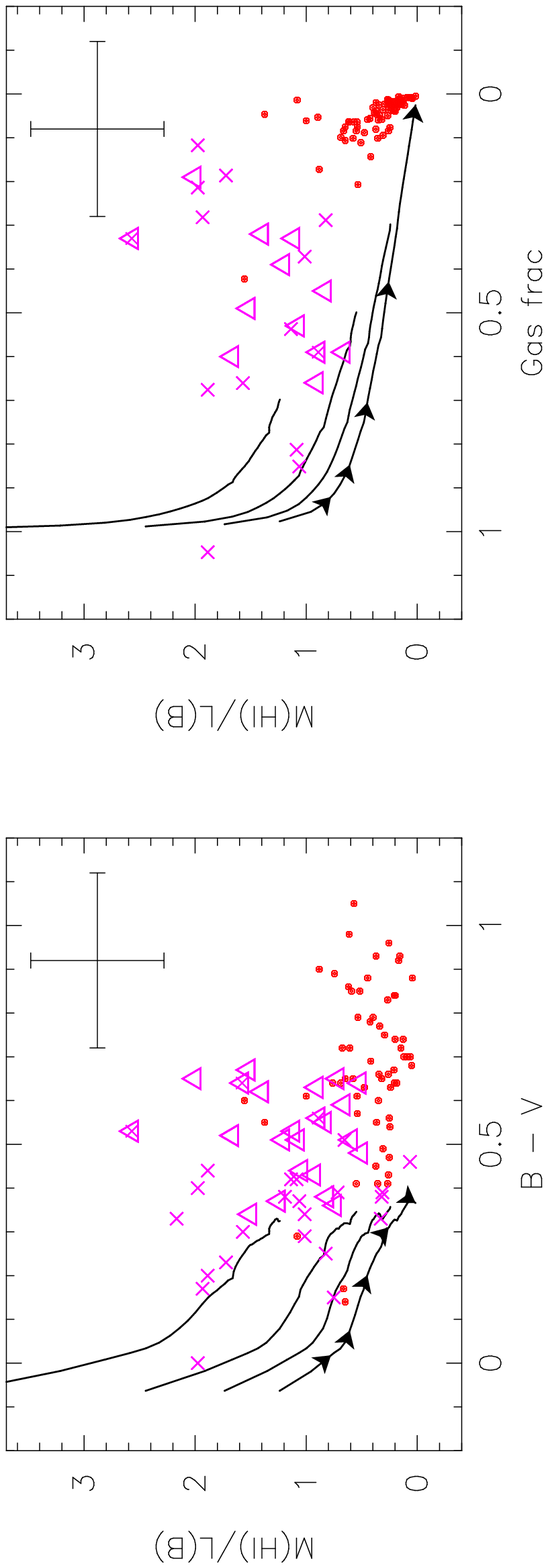},height=7.cm,width=16.cm,angle=270.}}
\caption[]{Evolution of $M_{{\rm HI}}/L_{B}$ {\it vs} $(B-V)$ (left) 
and $M_{{\rm HI}}/L_{B}$ {\it vs} $\mu_{1}$ (right). {\em Top panels:}
exponentially decreasing SFR models, {\em bottom panels:} constant SFR
models. From bottom to top, curves are shown for models ending at
increasing gas fractions $\mu_{1}  =$ 0.1, 0.3, 0.5, and 0.7,
respectively. Arrows on top of the $\mu_{1} = 0.1$ model indicate
evolution times of 1, 2, 4, 8, and 14 Gyr, respectively. Observational
data as in Fig.~8}
\end{figure*}

\begin{figure} 
\centerline{{\psfig{figure=\lsbfil{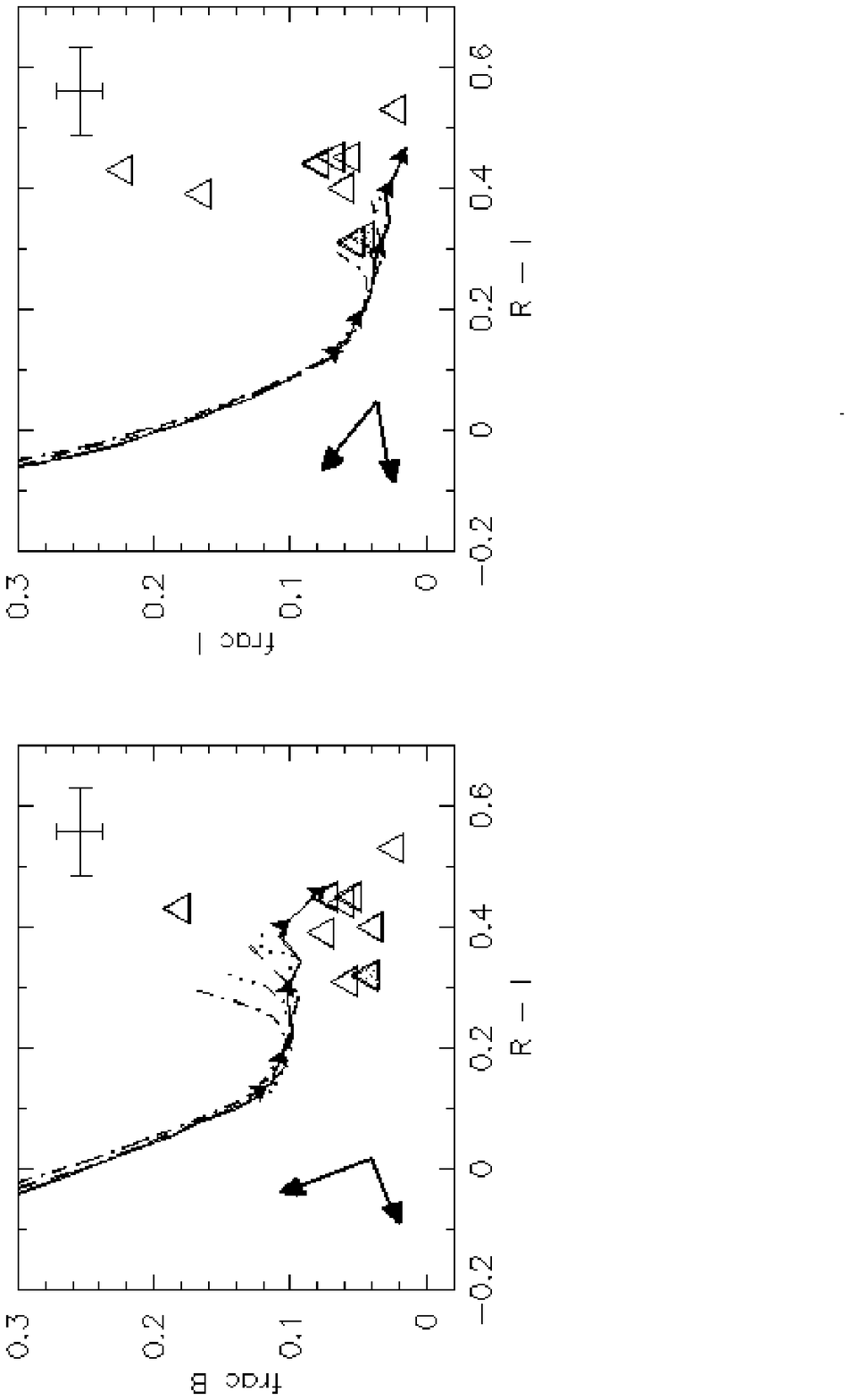},clip=,angle=270,height=6.cm,width=7.cm}}}
\caption[]{Evolution of the H{\sc ii} region integrated luminosity 
contribution for different SFR models ending at $\mu_{1} = 0.1$.
Shown is the ratio of H{\sc ii} region integrated and total luminosity
in the $I$ band $\eta_I$ {\it vs} $(R-I)$. For the exponentially
decreasing SFR model, solid triangles indicate evolution times of 1,
2, 4, 8, and 14 Gyr, respectively.  Observational data on LSB galaxies
is shown as open triangles. Typical errors in the data (upper right)
and directions to which the data would shift after corrections for
extinction (bottom left) are indicated.  Top and bottom arrows
indicate corrections for extinction within the H{\sc ii} regions and
the LSB galaxy for $E_{R-I} = 0.1$, respectively}
\end{figure}

\begin{figure}[thb]
{\psfig{figure=\lsbfil{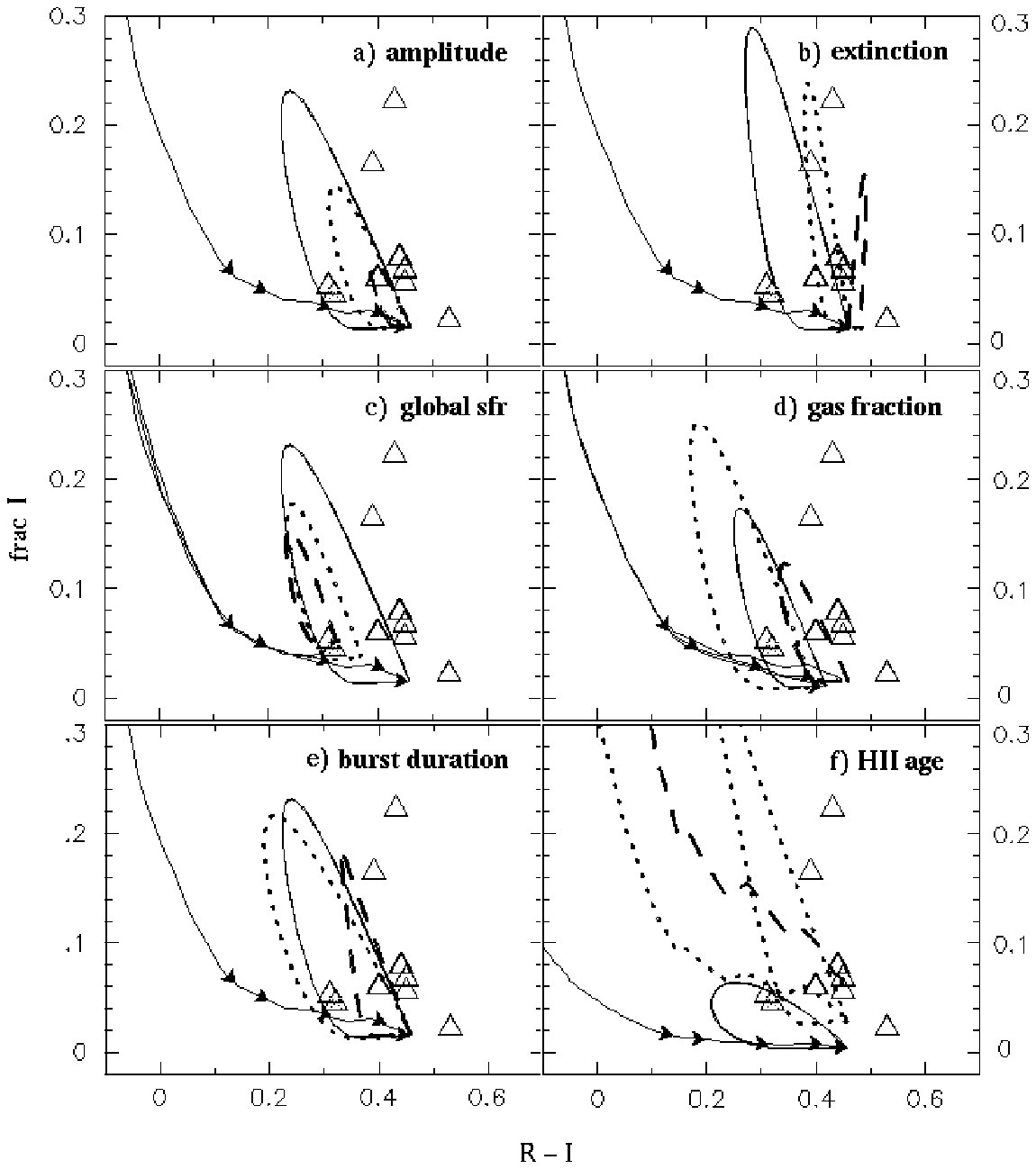},clip=,height=11.cm,width=9.cm,angle=0.}}
\caption[]{Impact of small amplitude bursts on the 
evolution of the H{\sc ii} region integrated luminosity contribution
in the $I$-band $\eta_I$ {\it vs} $(R-I)$. Unless stated otherwise, we plot the
exponentially decreasing SFR + burst model assuming $A_{\rm b} = 8$
\ms\, yr$^{-1}$, $\Delta t_{\rm b} = 5$ Myr, $\tau_{{\rm HII}} = 5$
Myr, and $E_{{B}-{V}} = 0$ mag.
{\bf a)} effect of varying burst amplitude: $A_{\rm b} = 8$ 
(solid curve), 4 (dotted), and 1.6 (dashed) \ms\, yr$^{-1}$;
{\bf b)} effect of mean extinction within H{\sc ii} regions:
$E_{{B}-{V}}$ = 0.25 (solid curve), 0.5 (dotted), and 1 
(dashed);
{\bf c)} effect of global star formation history: exponentially 
decreasing SFR (solid curve), constant (dotted), and linearly 
increasing (dashed);
{\bf d)} effect of actual gas-to-total mass ratio: $\mu_{1} = 
0.1$ (dashed curve), 0.5 (solid), and 0.7 (dotted);
{\bf e)} effect of burst duration: $\Delta t_{\rm b}$  = 1 (dashed 
curve), 5 (solid), and 10 (dotted) Myr; 
{\bf f)} effect of H{\sc ii} region age: $\tau_{{\rm HII}}$ 
= 2 (solid curve), 10 (dotted), and 50 (dashed; no burst shown) Myr.
Symbols have the same meaning as in Fig.~10.}
\end{figure}

We show the result in Fig.~11a.  At $t=t_b=13$ Gyr the contribution by
young stars increases rapidly. Simultaneously, the $(R-I)$ colours
become significantly bluer. After burst maximum, the contribution by
young stars decreases again and galaxy colours start to redden until
the effect of the burst becomes negligible and colours and magnitudes
evolve as prior to the burst.  

In this manner, a characteristic burst loop is completed as shown in
Fig.~11a. The shape of this loop is determined by: i) the burst
amplitude, ii) the extinction within the H{\sc ii} regions, iii) the
contribution by the old stellar population to the integrated galaxy
light, iv) the duration (and profile) of the burst, v) the maximum
time { during which young stars produced by the H{\sc ii} regions
can still be distinguished from the surrounding field population.}

\noindent {\bf Burst amplitude and IMF:} Fig.~11a 
demonstrates that values of $\eta_{I} \sim 0.25$ can be well explained
by bursts with amplitude $A_{\rm b} = 8$ \ms yr$^{-1}$ superimposed on
an exponentially decreasing SFR model ending at $\mu_{1} =
0.1$. Decreasing the burst amplitude by factors 2.5 and 5,
respectively, results in the smaller loops shown in Fig.~11a. The
actual burst amplitude required to explain the observations depends on
many quantities as described below

\noindent {\bf Dust extinction:} Fig.~11b illustrates 
the effect of dust extinction on $\eta_I$ and $(R-I)$.  A selective
extinction of $E_{{B}-{V}} = (0.25, 0.5, 1)$ mag for the H{\sc ii}
regions results in a reduction $\eta_I$ by a factor (1.6, 2.5, 6.5)
and a reddening $E_{R-I} =$ (0.14, 0.27, 0.56) mag, assuming a
Galactic extinction curve. For values of $E_{{B}-{V}} \ga 0.5$ mag, we
find that the bluing effect on $(R-I)$ by young massive stars formed
during the burst is neutralised almost entirely by extinction. For
intense bursts reddening of the galaxy colours may even occur.  If
variations in the mean extinction of the ensembles of H{\sc ii}
regions among LSB galaxies are small (\eg less than a factor two), it
is difficult to see how extinction alone can provide an adequate
explanation for the large variations observed in $\eta_{I}$.

\noindent {\bf Global star formation history:} 
From Fig.\ 11c it is clear that the burst effect on the galaxy
magnitudes and colours increases when the contribution by the old
stellar population is decreased. Thus, colours and magnitudes are less
affected by bursts imposed on constant or even increasing SFRs
compared to those imposed on exponentially decreasing SFR models: the
smaller loop sizes just reflect that the mean age of the underlying
stellar population is relatively low. 

\noindent {\bf Current gas fraction:} Fig. 11d demonstrates how the  
present-day gas fraction $\mu_{1}$ affects the effect of the burst.
The luminosity contribution by the old stellar population decreases
for increasing values of $\mu_{1}$, and thus, as discussed above, we
find that the effect of a star formation burst for galaxies with
$\mu_{1} = 0.9$ (\ie unevolved systems) is as large as that of a ten
times stronger burst for $\mu_{1} = 0.1$ (\ie highly evolved
systems). Thus, the burst amplitude required to explain the
observations strongly depends on the present-day gas-to-total
mass-ratio.  

\noindent {\bf Burst duration:} Fig.~11e shows that the duration of the 
burst affects the impact of the burst as well. For $\Delta t_{\rm b} =
1$, 5, and 10 Myr, the variation in $\eta_{I}$ is $\sim 0.25$ while
the resulting galaxy $(R-I)$ colours become bluer.  Burst durations in
excess of $\Delta t_{\rm b} \sim 5$ Myr are unlikely since this would
require dust extinctions $E_{(B-V)} \ga 1$ mag in order to provide
agreement with the observed $(R-I)$ colours (\cf Fig.~11b). Such large
extinction in H{\sc ii} regions are probably excluded by the
observations (\eg McGaugh 1994). $\bull \bull
\bull$ Thus, relatively narrow burst profiles are needed to explain
extreme values $\eta_{I} \sim 0.2$ (as in F568-V1; \cf Table~4). 

\noindent {\bf Maximum age of H{\sc ii} regions:} Fig.~11f 
shows that when the maximum lifetime of the H{\sc ii} regions is
increased from $\tau_{{\rm HII}} = 5$ to 50 Myr, partial agreement
with the observations can be achieved {\em without} invoking star
formation bursts. 
We note that the resulting H{\sc ii} region
contributions do not increase linearly with $\tau_{{\rm HII}}$ as
short lived massive stars dominate the luminosity contribution of all
stars formed during the past $\tau_{{\rm HII}}$ yr.

Values of $\tau_{{\rm HII}} \ga 50$ Myr would mean that stars down to
masses of $m_{\rm ion} \la 7$ \ms\ would contribute to the H{\sc ii}
regions identified (\eg Schaerer \etal 1993; $Z=0.001$).
Observational estimates for $m_{\rm ion}$ are usually in the range
$10-15$ \ms\ (Wilcots 1994; Garc\'{\i}a-Vargas 1995) and
correspond to $\tau_{{\rm HII}} \ga 15$ Myr.  Even though these values
would imply that our adopted value of $\tau_{{\rm HII}} = 5$ Myr is
too low, this excludes extreme values of $\tau_{{\rm HII}} = 200$ Myr
which would be required to explain the observed range in $\eta_{I}$
exclusively in terms of variations in $\tau_{{\rm HII}}$ and/or
$m_{\rm ion}$. Therefore, variations in $\tau_{{\rm HII}}$ (or
equivalently $m_{\rm ion}$) and/or extinction may provide an
explanation only for part of the variations in the H{\sc ii} region
integrated luminosity contributions observed among LSB galaxies.  \\


\begin{table*}
\caption[]{Maximum changes induced on galaxy magnitudes and colours for $M_g = 10^{10}\ M_{\odot}$\\ by a 5 Myr starburst}
\begin{tabular}{lclllccl}
\hline
Model & $A_{\rm b}$ & $\Delta B$ & $\Delta I$ & $\Delta(B-V)$ & 
 $\Delta(R-I)$ & Notes \\
& [\ms\, yr$^{-1}$] & [mag] & [mag] & [mag] & [mag] & \\ \hline
A1+burst & 0.8 & $-$0.34 & $-$0.14 & $-$0.18 & $-$0.04 & SFR$_{1}$ = 0.17 \ms 
 yr$^{-1}$ \\
"         & 1.6 & $-$0.58 & $-$0.30 & $-$0.28 & $-$0.10 & \\
"         & 3.0 & $-$1.05 & $-$0.42 & $-$0.43 & $-$0.17 & {\bf *} \\
"         & 8.0 & $-$1.72 & $-$0.53 & $-$0.56 & $-$0.26 & {\bf *} \\
B1+burst & 0.8 & $-$0.14 & $-$0.06 & $-$0.08 & $-$0.02 & SFR$_{1}$ = 0.89 \ms 
 yr$^{-1}$ \\
"         & 1.6 & $-$0.32 & $-$0.21 & $-$0.11 & $-$0.07 & \\
"         & 3.0 & $-$0.56 & $-$0.32 & $-$0.21 & $-$0.11 & {\bf *} \\
"         & 8.0 & $-$0.94 & $-$0.41 & $-$0.33 & $-$0.14 & {\bf *}
\\ \hline \\
\end{tabular}

{\bf *} $E_{(B-V)} \ga 0.3$ mag in H{\sc ii} regions is required to
provide agreement with observations
\end{table*}

In conclusion, the observations suggest that small amplitude, short
bursts of star formation are important in at least several of the LSB
galaxies for which accurate photometry data is available.  Such recent
episodes of enhanced star formation may play an important role in
affecting the colours of the blue LSB galaxies.

\subsection{Quantitative effect of small amplitude bursts on galaxy colours 
and magnitudes}

Table~5 lists the effect of a 5 Myr star formation burst on the galaxy
integrated magnitudes and colours for various burst amplitudes. For an
exponentially decreasing SFR (model A1 in Table~3; assuming $M_{\rm
g}(0) = 10^{10}$ \mss, $\mu_{1} = 0.1$, Salpeter IMF), we find that a
burst with amplitude $A_{\rm b} = 0.8$ \ms\, yr$^{-1}$ results in
maximum colour variations $\Delta(B-V)$ and $\Delta(R-I)$ of $-0.18$
and $-0.04$ mag, respectively.  The effect of increasing the burst
amplitude by a factor 10 to $A_{\rm b} = 8$ \ms\, yr$^{-1}$ results in
corresponding shifts of $-0.56$ and $-0.26$ mag, respectively. This
effect is similar to that when the initial galaxy mass is reduced by a
factor ten while leaving the burst amplitude unaltered (i.e.\ for.
$M_{\rm g}(0) = 10^{9}$ \mss\ and $A_{\rm b} = 0.8$ \ms\, yr$^{-1}$).

For bursts superimposed on exponentially decreasing SFRs models ending
at $\mu_{1} = 0.1$, the colour and magnitude shifts predicted are
consistent with the observations in case of burst amplitudes $A_{\rm
b} \la 3$ \ms\, yr$^{-1}$, assuming a typical extinction of $E_{(B-V)} =
0.3$ mag in the H{\sc ii} regions.  In fact, the effect of the burst
is determined mainly by the total luminosity of the young stellar
populations formed according to the continuous SFR during the last
Gyr or so.  Since the amplitude of the SFR scales with $(1- \mu_{1})$ for
models ending at different gas fractions $\mu_{1}$ the impact of the
burst for models ending at $\mu_{1} =  0.5$ is about twice that given
in Table~5. Similarly, for models ending at $\mu_{1} \ga 0.9$, the
burst effect becomes roughly ten times stronger compared to the
$\mu_{1} = 0.1$ case (\cf Fig.~11).
The effect of the burst is substantially reduced when going from
exponentially decreasing to constant SFRs (\cf Table~5).  

\begin{table}
\begin{minipage}{3.5cm}
\caption[]{Model SFR and gas fractions}
\begin{tabular}{lc}
\hline
(1) & (2) \\
$\mu_{1}$ &$A$ \\
& $(M_{\odot}{\rm yr}^{-1})$\\
\hline
0.1 & 0.18 \\
0.3 & 0.13 \\
0.5 & 0.09 \\
0.7 & 0.06 \\
0.9 & 0.02 \\
\hline
\end{tabular}
\end{minipage}
\end{table}

 
\begin{table*}
\caption[]{Observational estimates of present-day SFRs [\ms yr$^{-1}$] in LSB galaxies$^{*}$}
\begin{tabular}{lllllcllllll}
\hline
(1) & (2) & (3) & (4) & (5) & (6) & (7) & (8) & (9) & (10)  & (11) & (12) \\
Name & Dist. & $m_{I}$ & R$^{27}_{I}$ & R$_{HI}$ & 
$\mu_{I}$ & M$_{\rm HI}$ & 
$\langle \sigma_{\rm HI}\rangle$ & SFR$^{\rm fix}_{1}$ & SFR$_{1}$ & 
SFR$^{\rm cont}$ & SFR$^{\rm tot}$ \\
& [Mpc] & [mag] & [kpc] & [kpc] & [mag arcsec$^{-2}$] & [\ms] & 
[\ms pc$^{-2}$] &\multicolumn{4}{c}{[\ms yr$^{-1}$]}   \\ \hline
F561-1  & 47 & 14.7 &  6.6 &  7.6 & 23.2 & 8.91 & 4.5 & 0.05 & 0.09
& 0.013 & 0.081  \\
F563-1  & 34 & 16.1 & 10.2 & 16.3 & 26.4 & 9.19 & 1.0 & (0.03) & (0.02) & 
0.16 & 1.7\dag \\ 
F563-V1 & 38 & 15.8 &  4.6 &  4.8 & 24.0 & 8.48 & 4.2 & 0.02 & 0.03 & 
0.005 & 0.024 \\
F567-2  & 56 & 15.9 &  8.4 & 10.6 & 24.6 & 9.09 & 3.5 & 0.04 & 0.06 & 
0.029 & 0.15 \\ 
F568-1  & 64 & 15.0 &  9.6 & 11.5 & 23.7 & 9.35 & 5.4 & 0.08 & 0.18 &  
0.12 & 0.31 \\
F568-3  & 58 & 14.7 &  8.7 & 11.4 & 23.3 & 9.20 & 3.9 & 0.09 & 0.14 & 
0.06 & 0.33 \\ 
F568-V1 & 60 & 15.7 &  8.4 & 10.7 & 24.3 & 9.14 & 3.8 & 0.05 & 0.07 & 
0.05 & 0.58 \\
U128    & 48 & 13.5 & 18.2 & 21.4 & 24.1 & 9.55 & (2.0) & (0.16) & (0.21) & 
0.14 & 0.82 \\
U1230   & 40 & 14.3 & 12.0 & 18.8 & 24.5 & 9.51 & (4.3) & (0.11) & (0.15) & 
0.05 & 0.39 \\ \hline
\\
\end{tabular}

Notes: $^{*}$ theoretical values in columns (11) and (12) repeat those
in columns (9) and (11) from Table~4; {\bf \dag} probably contaminated
by field galaxy.  Values between parentheses are uncertain.

\end{table*}

\section{Present-day star formation rates in LSB galaxies}

\subsection{Theoretical star formation rates}

\subsubsection{Continuous SFR}
Theoretical SFRs can be derived from the
models discussed in the previous section (exponentially
decreasing SFR, Salpeter IMF), using:

\begin{equation}
{\rm SFR}^{\rm cont}  \approx A( \mu_{1} )  \frac{ M^{\rm
tot} }{10^{10}} \:\ \  {M}_{\odot}\, {\rm yr}^{-1}
\end{equation}
where $A(\mu_{\rm 1})$ is the model SFR amplitude required to end at
a gas fraction $\mu_{1}$ at a galactic evolution time of 14 Gyr
(assuming an initial mass of $10^{10}$ \mss) and $M^{\rm tot}$ the
total galaxy mass as obtained from $M_{{\rm HI}}$ and $\mu_{\rm 1}$.

We list the required values for $A$ to have the models end at certain
values of $\mu_{1}$ in Table~6.  Using these values we find that LSB
galaxies show present-day SFRs (without bursts) between $ 0.01 \la
{\rm SFR}^{\rm cont} \la 0.15 $\ms yr$^{-1}$.  For a typical LSB galaxy (\ie
$M^{\rm tot} = 10^{10}$ \mss, $\mu_{1} = 0.5$) we estimate SFR$^{\rm
cont} = 0.1$ \ms yr$^{-1}$.

\subsubsection{Burst SFR}

As we discussed in Sect.~6, the effect of a starburst on the galaxy
integrated magnitudes and colours depends on many quantities.
However, assuming a modest H{\sc ii} region extinction of
$E_{(B-V)} = 0.25$ mag and lifetime $\tau_{{\rm HII}} = 20$ Myr, crude
estimates for the maximum burst amplitude can be derived from:
\begin{equation}
{\rm SFR}^{\rm burst} \: \approx 20 \eta_{\rm I} 
\: \frac{M_{{\rm HI}}}{10^{10}} \: 
\left( \frac{1}{\mu_{\rm rot}} - 1 \right) \\ \ {M}_{\odot} {\rm yr}^{-1}
\end{equation}
We find that $I$-band contributions of $\eta_{\rm I} = 0.2$ are best
explained by a 5 Myr burst with amplitude SFR$^{\rm burst} = 0.8$ \ms
yr$^{-1}$, assuming present-day values of $M_{{\rm HI}} = 2\cdot
10^{9}$ \ms\ and $\mu_{\rm rot} = 0.5$.

Theoretical estimates for the total current SFRs in LSB galaxies are
found from ${\rm SFR}^{\rm tot} = {\rm SFR}^{\rm cont} + {\rm
SFR}^{\rm burst}$ and are listed in the last two columns of
Table~7. Present-day SFRs for LSB galaxies experiencing small
amplitude bursts range from SFR$^{\rm tot} \sim 0.02$ to 0.8 \ms\,
yr$^{-1}$.

\subsection{Empirical star formation rates}

We have also derived current SFRs for LSB galaxies using the empirical
method presented by Ryder \& Dopita (1994; hereafter RD) based on CCD
surface photometry of galactic disks.  These authors found a
relationship between the local H$\alpha$ and $I$-band surface
brightness in the disks of a sample of 34 of southern spiral galaxies.
From this relation, RD derived a constraint on the present-day SFR
integrated over the entire stellar mass spectrum as:
\vbox{\[
\log {\rm SFR_{1}} 
= -0.26 \mu_{I} + 0.92 \log \sigma_{\rm HI} + 5.3 
\]
\begin{equation}
\hfill {M}_{\odot} {\rm pc}^{-2} {\rm Gyr}^{-1} \ \ \ \ 
\end{equation}
}
where $\mu_{I}$ is the $I$-band surface brightness and $\sigma_{\rm
HI}$ is the global mean H{\sc i} surface density [\ms pc$^{-2}$]
within the star-forming disk.  The relation between SFR$_{1}$ and
$\mu_{I}$ is normalised by a term related to the mean surface density
$\sigma_{\rm HI}$ and by a constant which is partly related to the
conversion of the massive star formation rate to total SFR depending
on the adopted IMF (\cf Kennicutt 1983). It is unclear from the RD
sample whether the relation is valid also for the lowest (stellar)
$I$-band surface densities that are observed among LSB galaxies.
However, since this relation holds over a wide range in surface
brightness and massive star formation in the disks of spirals it appears
rather insensitive to galactic dynamics, extinction, and molecular gas
content (RD), we expect this relation to be valid also in case of LSB
galaxies.  At the faintest surface brightnesses (\ie $\mu_{I}
\ga 25.6$ mag arcsec$^{-2}$), the relation may be flattening off
although the effects of sky subtraction and small number statistics
leave this open to question. If flattening indeed occurs, Eq.~(5)
provides lower limits to the actual SFRs in LSB galaxies.

Using Eq.~(3), we estimate global present-day SFRs for all LSB
galaxies in our sample with measured $I$-band magnitudes and related
data.  For these LSB galaxies we list the distance, apparent $I$-band
magnitude, and radius $R_{27}$ of the 27 mag arcsec$^{-2}$ $B$-band
isophote, in columns (2) to (4) in Table~7. This radius corresponds to
the optical edge of the LSB galaxy and is more representative for the
radius within which the old disk stellar population in LSB galaxies is
contained than is $R_{25}$ as used by RD for HSB
galaxies. Accordingly, we define an effective I band surface
brightness as:
\begin{equation}
\mu_{I}^{\rm eff} =  m_{I} + 2.5 \log (\pi R_{27}^{2})
\end{equation}
and use this in Eq.~(5). We tabulate the outermost radius of the
measured H{\sc i} rotation curve $R_{{\rm HI}}$, effective $I$ band
surface brightness $\mu_{I}^{\rm eff}$, total H{\sc i} mass derived within
$R_{{\rm HI}}$, and the mean global surface H{\sc i} density
$\sigma_{{\rm HI}}$ in columns (5) to (8), respectively. 
{ For a consistent comparison between effective surface brightness and global H{\sc i} surface density,
one ought to measure them out to the same radius (e.g.\ $R_{27}$).}
However, since $M_{\rm HI}$ has been measured
within $R_{{\rm HI}}$, we expect the former values to be more
representative of the average H{\sc i} surface density in the star
forming part of the disk.
The mean global H{\sc i} surface densities $\sigma_{\rm HI}$ derived
using $R_{\rm HI}$ vary between 2 and 5.5 \ms pc$^{-2}$ and are
substantially smaller (\ie by $20-60$ \%) than those derived using
$R_{27}$ instead. 

The empirically derived mean present-day SFRs for the LSB galaxies
from our sample range from about SFR$_{1} = 0.02$ to $\sim 0.2$ \ms
yr$^{-1}$ (\cf column 10 of Table~7). Errors arising from the H{\sc i}
normalisation are estimated to be within a factor of two. This is
illustrated when the same SFRs are derived assuming a fixed H{\sc i}
surface density of 2 \ms pc$^{-2}$ for all LSB galaxies (\cf
SFR$_{1}^{\rm fix}$ in Table~7).

The empirically derived current SFRs in LSB galaxies are in good
agreement with the theoretically derived SFRs ranging from SFR$^{\rm
cont} \sim 0.01$ to 0.15 \ms yr$^{-1}$ (\cf Table~7). As discussed
before, the theoretically derived present-day SFRs of individual LSB
galaxies lie probably between SFR$^{\rm cont}$ and SFR$^{\rm tot}$
where the latter values include the contribution of small amplitude
bursts (SFR$^{\rm tot} =\ $SFR$^{\rm cont} + $SFR$^{\rm burst}$).  In
several cases, the values of SFR$^{\rm tot}$ are considerably larger
than the empirical values which suggests that the burst contribution
is overestimated by the models and/or that the SFRs derived
empirically do not trace well local enhancements of star formation at
the faint surface brightnesses of LSB galaxies.

The present-day SFRs in LSB galaxies are thus found to be considerably
below the $\sim 5-10$ \ms yr$^{-1}$ derived for their HSB counterparts
(\eg Kennicutt 1992 and references therein) but significantly larger
than the $\sim 0.001$ \ms yr$^{-1}$ observed typically in dwarf
irregular galaxies (\eg Hunter \& Gallagher 1986).

\section{Discussion}

In the previous sections we have focussed on the difference between HSB
galaxies and the type of LSB galaxies we have in our sample.  As noted
in the introduction (and also shown in Fig.~1) HSB and LSB galaxies,
however, do not form separate groups.  In fact there is a broad range of
galaxy properties and the purpose of the discussion is to clarify what
role the star formation history plays in determining the final character
of a galaxy. 
 
In general, there appears a trend along the Hubble sequence from rapidly
decaying SFRs for early type galaxies to constant or even increasing
SFRs for dwarf irregular galaxies (see \eg reviews by Sandage 1986;
Kennicutt 1992).  On average, the observed trend corresponds to a
decrease of the ratio of mean past to present SFR along the Hubble
sequence.  Most of the LSB galaxies in our sample belong to the group of
late-type galaxies for which exponentially decreasing SFR models
are in good agreement with the observations.  The remaining LSB
galaxies, for which slowly decreasing or constant (sporadic) SFR models
are more appropriate, belong to a group of galaxies with properties
intermediate to those of disk galaxies with weak or absent spiral arms
and Sm/Im galaxies.  Thus, in general LSB galaxies comply well with the
observed trend of SFR variation with Hubble type. 
 
The ``burst'' scenario discussed in section 6 indicates that current
star formation in virtually all the LSB galaxies in our sample is {\it
local} both in time and space and suggests that sporadic star formation
has been a continuous process from the time star formation started in the
disks of LSB galaxies. 
 
The low star formation rates of $\sim$0.1 \ms yr$^{-1}$ derived for LSB
galaxies as well as the local nature of the star formation in these
systems are consistent with the idea of a critical threshold for the
onset of global star formation in disk galaxies (\eg Skillman 1987;
Kennicutt 1989; Davies 1990). 
 
Even though LSB galaxies contain large amounts of gas, only very limited
amounts participate in the process of star formation.  If we assume that
LSB galaxies maintain their current star formation rate of $\sim$0.1 \ms
yr$^{-1}$, their typical present-day amount of gas $\sim$ M$_{\rm g}$ =
2.5 10$^{9}$ \ms\ will be consumed within $\tau_{\rm gas} \sim$30 Gyr
(for a recycled fraction of 25\%).  Such gas consumption times for LSB
galaxies are much larger than a Hubble time (\eg Romanishin 1980).  For
comparison, $\tau_{\rm gas} \sim 2-4$ Gyr in HSB galaxies, assuming
typically M$_{\rm g} \sim 10^{10}$ \ms\ and SFR$_{1} \sim 5$ \ms
yr$^{-1}$, which implies that HSB galaxies will run out of gas soon (see
Kennicutt 1992). 
 
The presence of an old stellar population in many late-type LSB
galaxies as indicated by their optical colours (e.g.  vdH93; dB95) and
as confirmed by the results from the modelling suggests that LSB
galaxies roughly follow the same evolutionary history as HSB galaxies,
{\it but at a much lower rate}.

The models suggest that very slowly decreasing star formation rates
with $e$-folding times much larger than $\sim$ 5 Gyr probably can be
ruled out for the majority of the LSB galaxies examined here.  The
observed colours are too blue, provided extinction is indeed
negligible.  This result is insensitive to the possible occurrence of
infall or accretion of gas.  Similarly the models indicate that the
dominant stellar population in galaxies with $(B - V) \ge 0.5$ mag
cannot be as young as 10 or 5 Gyr (cf.  Fig.~5).  In contrast the
models can not rule out a dominant luminosity contribution by stellar
populations significantly younger than 5 -- 10 Gyr for LSB galaxies
with $(B - V) \approx 0.4$.
 
First this indicates that the mean age of the stellar population in most
LSB and HSB galaxies is similar even though the disks of LSB galaxies
are in a relative early evolutionary stage.  Although we have not
explored an entire range of star formation e-folding times, values much
below or in excess of $\sim$ 5 Gyr are not very likely.  Smaller values
would increase the colour contribution from the underlying older stellar
population and hence produce redder colours than observed, while much
larger values correspond to slowly decreasing or almost constant star
formation rates implying a relatively larger contribution of the younger
stellar population. 

Secondly, the combined effect of extinction and metallicity on galaxy
colours is sufficient to explain the colour differences observed between
LSB galaxies and HSB galaxies.  Since the amount of extinction depends
strongly on the dust content, which in turn is coupled to the heavy
element abundances in the ISM (see Sect.  2), metallicity probably is
the main quantity determining the colour differences observed between
LSB and HSB galaxies.  In this manner the much lower rate of star
formation in LSB galaxies, implying lower metallicities and dust
contents, indirectly determines the blue colours of LSB galaxies
compared to HSB galaxies. 

\section{Summary}

We have examined the star formation histories of LSB galaxies using
models which take into account both the photometric and the chemical
evolution of the galaxies.
For the majority of the LSB galaxies in our sample, observed $UBVRI$
magnitudes, [O/H] abundances, gas masses and fractions, and H{\sc i}
mass-to-light ratios are best explained by an exponentially decreasing
global SFR ending at a present-day gas-to-total mass-ratio of $\mu_{1}
\sim 0.5$. When small amplitude bursts are involved to decrease the predicted
$M_{\rm g}/L$ ratios, models ending at $\mu_{1} \approx 0.7$ may also
apply. In addition to exponentially decreasing SFR models, $\sim 15\%$
of the LSB galaxies require modest amounts of internal extinction
$E_{(B-V)}\la 0.1$ mag to explain the relatively red colours of $(B-V)
\sim 0.6$ of these systems.

A substantial fraction ($\sim 35\%$) of the LSB galaxies in our sample
have colours $(B-V) \la 0.45$ mag and exhibit properties similar to
those resulting from exponentially decreasing SFR models at evolution
times of $\sim 5 - 10$ Gyr. Alternatively, recent episodes of enhanced
star formation superimposed on exponentially decreasing SFR models may
provide an adequate explanation for the colours of these systems (see
Sect. 6).  Recent star formation is observed, at least at low levels,
in essentially all the LSB galaxies in our sample. Hence, it seems
justified to assume that the disks in LSB galaxies experienced
continuous (\ie frequent small amplitude bursts of) star formation, at
least during the last few Gyr.

There is nothing peculiar about the evolution of LSB galaxies. Broadly
speaking their evolution proceeds like that of HSB galaxies, but at a
much lower rate.

\begin{acknowledgements}
We thank Andre Maeder and Georges Mey\-anet for providing us with the
stellar isochrone data and conversion programs.  We thank Roelof de
Jong for making available to us the data of a large sample of face-on
spirals.  We are grateful to the referee Uta Fritze-von Alvensleben
for constructive comments from which this paper has benefitted.
\end{acknowledgements}

\end{document}